\pdfoutput=1

\documentclass{JHEP3}

\usepackage{amsmath}
\usepackage{amssymb}
\usepackage{amsthm}
\usepackage{graphicx}
\usepackage{slashed}

\newcommand{\bC}{\mathbb{C}}
\newcommand{\bF}{\mathbb{F}}

\newcommand{\bP}{\mathbb{P}}
\newcommand{\bR}{\mathbb{R}}
\newcommand{\bZ}{\mathbb{Z}}
\newcommand{\cA}{\mathcal{A}}
\newcommand{\cB}{\mathcal{B}}
\newcommand{\cC}{\mathcal{C}}

\newcommand{\cF}{\mathcal{F}}
\newcommand{\cI}{\mathcal{I}}
\newcommand{\cL}{\mathcal{L}}
\newcommand{\cM}{\mathcal{M}}
\newcommand{\cN}{\mathcal{N}}
\newcommand{\cO}{\mathcal{O}}
\newcommand{\cP}{\mathcal{P}}
\newcommand{\cR}{\mathcal{R}}
\newcommand{\cS}{\mathcal{S}}

\newcommand{\fC}{\mathfrak{C}}

\newcommand{\fP}{\mathfrak{P}}
\newcommand{\fS}{\mathfrak{S}}
\newcommand{\Tr}{\mathrm{Tr\,}}
\newcommand{\ov}{\overline}

\newtheorem*{thm}{Theorem}

\newtheorem*{conjecture}{Conjecture}
\newtheorem*{solvability}{Solvability Conditions}

\DeclareMathOperator{\ch}{ch}
\DeclareMathOperator{\rk}{rk}
\DeclareMathOperator{\Td}{Td}

\title{\centering{On the computation of non-perturbative effective potentials in
  the string theory landscape\\ \vspace{.2cm}  \Large --- IIB/F-theory perspective ---}}
\author{Mirjam Cveti\v{c}$^{1,2,3}$, I\~naki Garc\'ia-Etxebarria$^{1,2}$ and James Halverson$^{1,2}$\\
  $^1$Department of Physics and Astronomy, University of Pennsylvania,\\
  Philadelphia, PA 19104-6396, USA\vspace{.2cm}\\
  $^2$Kavli Institute for Theoretical Physics, Kohn Hall,\\
  UCSB, Santa Barbara, CA 93106, USA \vspace{.2cm}\\
  $^3$Center for Applied Mathematics and Theoretical Physics,\\
  University of Maribor, Maribor, Slovenia\vspace{.5cm} \\
  E-mail: \email{cvetic@cvetic.hep.upenn.edu},
  \email{inaki@sas.upenn.edu}, \email{jhal@physics.upenn.edu}}

\abstract{We discuss a number of issues arising when computing
  non-perturbative effects systematically across the string theory
  landscape. In particular, we cast the study of fairly generic
  physical properties into the language of computability/number theory
  and show that this amounts to solving systems of diophantine
  equations. In analogy to the negative solution to Hilbert's 10th
  problem, we argue that in such systematic studies there may be no
  algorithm by which one can determine all physical effects.  We take
  large volume type IIB compactifications as an example, with the
  physical property of interest being the low-energy non-perturbative
  F-terms of a generic compactification. A similar analysis is
  expected to hold for other kinds of string vacua, and we discuss in
  particular the extension of our ideas to F-theory. While these
  results imply that it may not be possible to systematically answer
  certain physical questions about generic type IIB compactifications,
  we identify particular Calabi-Yau manifolds in which the diophantine
  equations become linear, and thus can be systematically
  solved. \vspace{.5cm}

  As part of the study of the required systematics of F-terms, we
  develop technology for computing $\bZ_2$ equivariant line bundle
  cohomology on toric varieties, which determines the presence of
  particular instanton zero modes via the Koszul complex. This is of
  general interest for realistic IIB model building on complete
  intersections in toric ambient spaces.}

\preprint{UPR-1219-T\\NSF-KITP-10-124}

\begin{document}

\section{Introduction}

The string theory landscape is vast and unwieldy. In order to
understand its dynamics from a top-down perspective, we would like to
have a method of constructing the low energy effective action given
the compactification data for a string vacuum. In this paper we deal
with the question of whether such a method exists using results in
computability theory. More precisely, we will describe the
computability structure of a well-defined subproblem of the problem of
computing low energy effective actions, namely the problem of
determining the non-perturbative part of F-terms for all vacua.

There are various reasons to be interested in computing
non-perturbative F-terms when building realistic string models. Most
prominently, in the context of IIB/F-theory compactifications that we
study in this paper, D-brane instantons are the only effect that can
possibly lift some directions in the classical moduli space that
remain flat to all loop orders. The importance of D-brane instantons
in the context of K\"ahler moduli stabilization is well known
\cite{Denef:2004dm,Balasubramanian:2005zx,Denef:2005mm}, and recently
it has been realized that the picture is significantly enriched when
couplings of instantons to charged zero modes is included in the
discussion
\cite{Blumenhagen:2006xt,Ibanez:2006da,Florea:2006si,Blumenhagen:2007sm}
(for a review see \cite{Blumenhagen:2009qh}). In particular, couplings
in the open string sector that are forbidden in perturbation theory do
arise when one takes into account the effect of D-brane
instantons. There is another reason why we want to focus on D-brane
instantons in this paper. As we will see below, the computability
structure of D-brane instantons can be made particularly clear in the
context of IIB/F-theory backgrounds, and some interesting links to
well known results in number theory arise when looking at things from
this perspective.

In particular configurations, exact methods for computing
non-perturbative F-terms exist. For example, the low-energy
prepotential including all instanton corrections for many $\cN=2$
field theories in four dimensions has been obtained exactly using the
techniques in \cite{Seiberg:1994rs,Seiberg:1994aj}. This can even be
reformulated directly in terms of instanton counting
\cite{Nekrasov:2002qd}.

Realistic particle physics theories have at most $\cN=1$ supersymmetry
in four dimensions, and here the knowledge is more limited. Some
$\cN=1$ theories admit a matrix model description, and can be solved
exactly using the techniques in
\cite{Dijkgraaf:2002fc,Dijkgraaf:2002vw,Dijkgraaf:2002dh}. This exact
solution also sheds light on the problem of classifying instantons
contributing to different F-terms
\cite{Aganagic:2007py,GarciaEtxebarria:2008iw}, a problem that we will
discuss extensively below. One can also approach the problem in other
ways. For example, one can try to translate known results coming from
the topological string to the physical $\cN=1$ theory
\cite{Collinucci:2009nv,Petersson:2010qu}, directly orientifold or add
fluxes to known $\cN=2$ results
\cite{Vafa:2000wi,Lawrence:2004zk,Davidse:2005ef,Saueressig:2005es,Alexandrov:2010np,Camara:2008zk},
or, in cases where enough of the S-duality group survives, one can use
it to constrain the form of the superpotential \cite{Grimm:2007xm}.

The cases where the low energy physics can currently be derived
exactly are nevertheless rather special, and one may wonder if
anything can be said about the problem in generality, that is, across the
whole landscape of string vacua. In this paper we argue that the
structure of vacua in the landscape determines whether the problem of
determining exact F-terms systematically has a solution or not. The
more generic the number theoretical properties of string
vacua,\footnote{We define carefully what we mean by this in
  section~\ref{sec:hilbert}.} the more likely it is that there is no
generic method for computing the low energy effective field theory for
an arbitrary compactification.

In order to show this fact, we show that existence of a method for
systematically computing the effective field theory for a
compactification implies the existence of a method for determining
whether a particular class of integer (diophantine) equations derived
from the geometric compactification data are solvable. For completely
generic diophantine equations, this is known to be impossible to do
algorithmically. So we conclude that either:
\begin{itemize}
\item The space of possible $\cN=1$ vacua of string theory has some
  hidden algebraic structure, such that the low energy physics can be
  solved algorithmically, or
\item The low energy superpotential for a generic string
  compactification cannot be computed.
\end{itemize}
To us, the second option seems to be more likely (below we give
arguments why), although the first one is also a very interesting
possibility, if true, and at the moment we have no way of determining
which one is actually realized. In order to give concrete evidence for
this result, we will work in a particularly tractable corner of the
landscape, namely large volume type IIB compactifications.

Our discussion is organized as follows. In
section~\ref{sec:systematics} we will describe the general systematics
involved in answering questions about non-perturbative F-terms in
large volume type IIB compactifications and their lifts to
F-theory. In section~\ref{sec:hilbert} we will show a formalization of
the approach in section~\ref{sec:systematics} which illuminates the number
theoretical structure of the problem (in any approach, not necessarily
the one discussed in section~\ref{sec:systematics}). In this way we
formulate a natural conjecture for the
computability structure of the landscape based on known results for
very similar problems in number theory. In
section~\ref{sec:elliptic-fibrations} we turn the discussion around,
showing that a certain class of type IIB Calabi-Yau compactifications
admits an algorithm for computing a subset of F-terms
systematically. In particular we will see that threefolds with a
factorizing intersection form, which include many elliptically fibered
Calabi-Yau threefolds, have especially nice properties when it comes
to computing superpotentials.

Many of the tools we use in formulating the problem are known (this is
the main reason why we chose large volume IIB in the first place), but
in order to be completely explicit in our specific computations we
needed to develop efficient tools for computing $\bZ_2$ equivariant
line bundle cohomology on toric varieties. We explain how to do this
is in appendix~\ref{sec:equivariant-cohomology}.  Finally,
appendix~\ref{sec:geometric-viewpoint} contains a reformulation of
part of the discussion in section~\ref{sec:elliptic-fibrations} in
terms of lattice data, which may be illuminating to readers familiar
with toric geometry.

\section{Systematics}
\label{sec:systematics}

In this section we will review some basic tools for analyzing D-brane
instanton effects in type IIB compactifications, and in particular we
present the sheaf cohomology groups and geometric indices relevant to
instanton zero modes. The picture seems to be qualitatively similar in
F-theory, and we elaborate on this below. Our emphasis will be on
showing the existence of explicit computational methods for
determining the zero modes on the instanton. As we will see, in the
context of IIB/F-theory compactifications coming from a complete
intersection in toric varieties there is a systematic method for
computing the spectrum of instanton zero modes which can be completed
in finite time. The existence of this method will be important in
section~\ref{sec:hilbert}. For an extensive review of D-brane
instantons in type II string theory, see \cite{Blumenhagen:2009qh}.

We will be mostly dealing with euclidean D3-branes in type IIB
orientifolds. In particular, we will discuss configurations with
$O7^-$ planes, focusing on BPS instantons coming from euclidean D3
branes wrapping holomorphic divisors of the Calabi-Yau. The
zero modes of a euclidean D3 brane in such a background naturally
split into neutral modes and modes charged under the background
space-filling D7 branes. Neutral zero modes are those that do not have
charge under the D7 gauge group, while charged modes transform in the
(anti)fundamental representation. They arise from open strings going
from the instanton to itself and open strings going from the instanton
to the D7, respectively. Let us study each class in turn.

\subsection{Neutral zero modes}
\label{sec:neutral-modes}

Each BPS D-brane instanton is 1/2 BPS, and thus locally breaks 4 out
of the 8 supersymmetries of the background Calabi-Yau geometry. These
broken supersymmetries manifest themselves in the volume of the
instanton as Goldstinos, that is, as fermionic zero modes. These are
conventionally denoted as $\theta^\alpha$ and
$\ov\tau_{\dot\alpha}$. Depending on the divisor wrapped by the
instanton, some other neutral zero modes, such as deformation modes,
may be present.

\TABLE{
  \label{table:O(1)-modes}
  \begin{tabular}{c|c}
    Zero modes & Number\\
    \hline
    $(X_\mu,\theta^\alpha)$ & $h^0_+(D,\cO_D)=1$\\
    $\ov\tau_{\dot\alpha}$ & $h^0_-(D,\cO_D)=0$\\
    $\gamma_\alpha$ & $h^1_+(D,\cO_D)$\\
    $(\omega,\ov\gamma_{\dot\alpha})$ & $h^1_-(D,\cO_D)$\\
    $\chi_\alpha$ & $h^2_+(D,\cO_D)$\\
    $(c,\ov\chi_{\dot\alpha})$ & $h^2_-(D,\cO_D)$
  \end{tabular}

  \caption{Zero mode structure for an $O(1)$ instanton wrapping a
    connected cycle $D$. We follow the conventions in
    \cite{Blumenhagen:2009qh,Blumenhagen:2010ja}. Modes with a spinor
    index are fermionic. The $\pm$ subindex denotes the parity with
    respect to the orientifold action.}
} In a typical $\cN=1$ large volume type IIB compactification there
are, in addition to the geometric Calabi-Yau background, various other
ingredients such as branes, orientifolds and fluxes. Some of these
ingredients will change the spectrum of zero modes. For instance,
instantons on top of branes have some of their zero modes lifted by
ADHM couplings. Or, if an instanton is mapped to itself under the
orientifold involution in such a way that only an $O(1)$ gauge group
survives, the $\ov\tau_{\dot\alpha}$ modes are projected
out. Similarly, background closed string fluxes can lift some neutral
zero modes of the instanton (see
\cite{GarciaEtxebarria:2008pi,Billo':2008sp,Billo':2008pg,Uranga:2008nh}
for recent works in this direction). In this paper we will ignore this
last possibility, and for our explicit examples we will construct
backgrounds without flux. The general structure for the neutral zero
modes of an $O(1)$ instanton was discussed in
\cite{Billo:2002hm,Ibanez:2006da,Argurio:2006ny,Argurio:2007qk,Argurio:2007vqa,Ibanez:2007rs,Blumenhagen:2007bn,Blumenhagen:2009qh,Blumenhagen:2010ja},
and we reproduce it in table~\ref{table:O(1)-modes}.

\medskip

If one is interested in contributions to the superpotential
of an $\cN=1$ compactification, then the instanton will
contribute if it has only the two $\theta^\alpha$ zero modes. In this
paper we will be interested in the more general situation of computing
all F-terms, so instantons with extra zero modes are acceptable and
interesting, giving rise to F-terms in the $\cN=1$ action with a
larger number of fermionic operators than two
\cite{Beasley:2004ys,Beasley:2005iu}. In some cases these effects can
be rather dramatic, for example the quantum deformation of the moduli
space of $N_f=N_c$ SQCD can be attributed to an instanton with four
fermionic zero modes \cite{Beasley:2004ys}. 

Nevertheless, the superpotential case is still particularly
interesting, so let us elaborate on some index techniques which are
useful in this case. Looking to table~\ref{table:O(1)-modes}, and
recalling that $H^i(D,\cO_D)=H^i_+(D,\cO_D)\oplus H^i_-(D,\cO_D)$, we
have that a necessary condition for a $O(1)$ instanton to contribute
to the superpotential is that:
\begin{align}
  \chi(D,\cO_D) = \sum_{i=0}^2 (-1)^i (h^i_+(D,\cO_D)+ h^i_-(D,\cO_D)) = 1.
\end{align}
This is easy to compute using the Riemann-Roch formula:
\begin{align}
  \label{eq:GRR}
  \chi(D,\cO_D) = \int_D \ch(\cO_D)\Td(TD) = \int_D \Td(TD)
\end{align}
where $TD$ is the tangent bundle to $D$. In
section~\ref{sec:elliptic-fibrations} we will make abundant use of
this formula.

Similarly, we also need to have that
\begin{align}
  \label{eq:Lefschetz-condition}
  \chi^\sigma(D,\cO_D) = \sum_{i=0}^2 (-1)^i (h^i_+(D,\cO_D)-
  h^i_-(D,\cO_D)) = 1.
\end{align}
where $\chi^\sigma$ denotes Lefschetz's equivariant genus for the
orientifold action $\sigma$. We give a brief introduction to the
equivariant genus in appendix~\ref{sec:equivariant-cohomology}.

An important observation, which helps motivate the abstract study
of diophantine equations in section~\ref{sec:hilbert}, is that both
\eqref{eq:GRR} and \eqref{eq:Lefschetz-condition} give rise to
equations for integer unknowns with integer coefficients. More
precisely, let us take a basis $\{D_i\}$ of divisors of the
Calabi-Yau, and write the cycle wrapped by the instanton as $D=\sum
d_i D_i$, with the $d_i$ being integers. Then~\eqref{eq:GRR} gives an
inhomogeneous equation of degree 3 on the $d_i$:
\begin{align}
  \chi(D,\cO_D) = a_{ijk} d_i d_j d_k + b_{ij} d_i d_j + c_i d_i = 1 
\end{align}
with the coefficients $a_{ijk},b_{ij},c_i$ being rational numbers
depending on the geometry (we can immediately obtain an equation with
integer coefficients by multiplying both sides of the equation by an
appropriate integer). Similarly the equivariant index
constraint~\eqref{eq:Lefschetz-condition} gives an inhomogeneous
equation of degree 2 on the $d_i$:
\begin{align}
  \chi^\sigma(D,\cO_D) = e_{ij} d_i d_j + f = 1
\end{align}
In general, if one is interested in computing all instantons
satisfying these conditions one is, in effect, providing a way of
solving this system of coupled diophantine equations. We stress that,
as we discuss extensively in section~\ref{sec:hilbert}, the
equivalence between questions about non-perturbative dynamics and
systems of diophantine equations runs much deeper than what this
straightforward discussion would suggest.

We also want to emphasize that, although all line bundle cohomology
and index calculations performed in the paper are done in the context
of toric geometry, all of the formulas for indices and sheaf
cohomology groups given in this section are not in any way specific to
Calabi-Yau manifolds that are complete intersections in toric
varieties. We choose this type of Calabi-Yau because calculations are
particularly tractable in toric geometry, and we can be precise about
the structure of the required computation.

\subsection*{Concrete computation of sheaf cohomology}

We see that the problem of computing the spectrum of neutral zero
modes on the instanton reduces to computing equivariant line bundle
cohomology. Recently work has been done and computer implementations
have been provided
\cite{Cvetic:2010rq,Blumenhagen:2010pv,Jow:2010,Rahn:2010fm} which
allow for the direct calculation of the (non-equivariant) sheaf
cohomology groups when the Calabi-Yau is a hypersurface or complete
intersection in an ambient toric variety. The basic procedure is to
first calculate relevant line bundle cohomology on the ambient space
which then determines $H^i(D,\cO_D)$ via the long exact sequences in
cohomology associated with Koszul sequences. It is not very hard to
extend this procedure to equivariant cohomology, which we show how to do
in appendix~\ref{sec:equivariant-cohomology}.

While it is always possible to algorithmically perform this
computation for a given divisor $D$, the problem of systematically
determining all instantons which might contribute to the
superpotential requires knowledge of $h^i_\pm(D,\cO_D)$ for a divisor
$D=\sum d_iD_i$ as an explicit function of the $d_i$. For simple
examples, one can express this function as a simple polynomial in the
$d_i$, as done in (for example) \cite{Blumenhagen:2010ja}, but for
generic backgrounds this will not be possible. This is due to the fact
that calculating line bundle cohomology on the ambient space amounts
to counting points in polyhedra in the $M$ lattice of the toric
variety, which can be solved analytically for simple polyhedra, but
not for generic polyhedra that arise in more complicated toric
varieties. The only available representation of the result is then in
terms of the algorithm that computes line bundle cohomology. This fact
is ultimately the reason why, in order to understand the generic
structure of non-perturbative F-terms in string theory, we are lead
into the mathematical theory of computability.

\subsection{Charged zero modes}
\label{sec:charged-modes}

Starting with \cite{Blumenhagen:2006xt,Ibanez:2006da,Florea:2006si} it
was realized that D-brane instantons provide a way of obtaining
couplings in semi-realistic models that are forbidden in perturbation
theory. Typical examples are $\mu$ terms \cite{Blumenhagen:2006xt},
neutrino Majorana masses
\cite{Blumenhagen:2006xt,Ibanez:2006da,Cvetic:2007ku,Ibanez:2007rs,Antusch:2007jd},
and $10\,10\,5_H$ couplings \cite{Blumenhagen:2007zk} in $SU(5)$ GUT
models. For a recent systematic study of instanton effects on
realistic type II MSSM quivers see
\cite{Cvetic:2009yh,Cvetic:2009ez,Cvetic:2009ng,Cvetic:2010mm} and
references therein.

The important observation is that, in addition to the open strings
going from the instanton to itself, there are also open strings going
from the instanton to the background D7 branes. These open strings
give rise to zero modes of the instanton transforming in the
fundamental or antifundamental of the D7 brane gauge group. These zero
modes, in turn, can couple to matter fields on the D7 branes, and when
integrating over the instanton zero mode measure this process can
induce effective operators involving matter fields in the low energy
effective action of the compactification.

This implies that in order to understand the physics of charged zero
modes we need to understand the structure of Yukawa couplings in our
compactification. In this paper we will sidestep this complication by
restricting our explicit examples in
section~\ref{sec:elliptic-fibrations} to questions that require the
absence of charged zero modes on the instanton. In the general
discussion of section~\ref{sec:hilbert} such a restriction is not
made.

In any case, the spectrum of zero modes is obtained easily as
follows. Consider a background D7 brane $A$ and an instanton $D$ such
that they wrap different divisors. They intersect over the curve $\cC
= A\cdot D$. The spectrum of zero modes between the instanton and the
brane then comes from the cohomology groups
\cite{Sharpe:2003dr,Cvetic:2007qj}:
\begin{align}
  (\alpha, \bar\beta) \in (H^0(\cC, K_\cC^{1/2}), H^1(\cC,
  K_\cC^{1/2}))
\end{align}
where $\alpha$ and $\ov\beta$ stand for the modes in the fundamental
and antifundamental respectively of the D7 brane group, and $K_\cC$
denotes the anticanonical bundle of $\cC$. The appearance of the
anticanonical class in these formulas comes ultimately from the fact
that branes are classified by K-theory, instead of just cohomology
\cite{Sharpe:2003dr}. In the examples in
section~\ref{sec:elliptic-fibrations} we will ensure that there are no
zero modes by ensuring that either
\begin{itemize}
\item $\cC=\bP^1$ and $\chi(\cC, K_\cC^{1/2}) = 0$, or alternatively,
\item $\cC=0$, i.e., there is no intersection.
\end{itemize}
As we will see in examples below, these conditions give respectively
quadratic and linear diophantine equations in the integers $d_i$ which
parameterize the divisor $D$ which the instanton wraps.

Let us mention for completeness that one can also wrap the D-brane
instanton on a cycle $S$ also wrapped by a D7 brane. In this case the
formulas above need some modification. The relevant formulas for the
charged zero modes between the instanton and the brane are
\cite{Heckman:2008es,Marsano:2008py}:
\begin{align}
  \begin{split}
    \alpha &\in H^0(S,K_S\otimes \cL)\oplus H^1(S,\cL)\oplus
    H^2(S,K_S\otimes \cL)\\
    \ov\beta &\in H^0(S,K_S\otimes \cL^*)\oplus H^1(S,\cL^*)\oplus
    H^2(S,K_S\otimes \cL^*)
  \end{split}
\end{align}
where $\cL$ is the line bundle on the D7 brane. For the applications
in the rest of the paper, these modes have little effect: since they
come from cohomology classes they are manifestly included in the
discussion in section~\ref{sec:hilbert} and, since they are a finite
number of instantons in this class that can possibly contribute to the
superpotential, they do not affect the discussion in
section~\ref{sec:elliptic-fibrations}. We will thus ignore this class
of instantons in the remainder of the paper.

\subsection{Worldvolume fluxes}
\label{sec:worldvolume-flux}

Although we will not consider this in the explicit examples in this
paper, let us mention that worldvolume fluxes are also easily included
in our discussion. Let us first discuss $O(1)$ instantons. In this
case the bundle on the instanton is essentially trivial, and we can
assume that it is in fact trivial for our discussion. The D7 brane can
still have a bundle $B$ on its worldvolume, which can give rise to
chirality between the brane and the instanton. More precisely, the
spectrum of charged zero modes is now given by:
\begin{align}
  \label{eq:charged-modes-no-L}
  (\alpha, \bar\beta) \in (H^0(\cC, B\otimes K_\cC^{1/2}), H^1(\cC,
  B\otimes K_\cC^{1/2}))
\end{align}

In the case of $U(1)$ instantons we can also consider putting a bundle
$\cL$ on the instanton. This introduces some new interesting features
into the problem. Notice that the problem now involves new integer
variables $l_i$, which determine the bundle $\cL$, in addition to the $d_i$.
For $\cL$ a line bundle we have that $\cL=\cO(\sum l_i D_i)$,
with $D_i$ some basis of divisors on the instanton worldvolume. In
scanning over all possible instantons in the given background we also
need to scan over the $l_i$. We may also need to make sure that the
instanton that we are considering is BPS, otherwise extra Goldstinos
will arise in the neutral sector, and the resulting coupling is most
naturally understood as a D-term.\footnote{Sometimes one is free to
  move in moduli space in such a way that the instanton becomes BPS at
  some point. In the BPS locus the extra Goldstinos turn into extra
  unsaturated fermionic zero modes, and the D-term is reinterpreted as
  a local description of a higher F-term
  \cite{GarciaEtxebarria:2007zv,GarciaEtxebarria:2008pi}.} This will
happen when the flux is \emph{primitive}:
\begin{align}
  \int_D c_1(\cL)\wedge(J+iB) = 0 
\end{align}
with $J$ and $B$ the K\"ahler form and $B$-field respectively. This
equation gives a quadratic equation involving $d_i$ and $l_i$:
\begin{align}
  p_{ij} d_il_j = 0.
\end{align}
This case is slightly different from the ones discussed so far, in that
even if $d_i,l_i$ are integer unknowns, the $p_{ij}$ coefficients are
general complex numbers, so we cannot talk of diophantine
equations. As we will discuss in section~\ref{sec:hilbert}, this
problem still admits a diophantine representation, as long as $J+iB$
is a \emph{computable} real number \cite{Turing36} (see
\cite{weihrauch2000} for a recent review of the field). Roughly
speaking, computable numbers are those numbers that can be computed
term by term in a decimal expansion to arbitrary precision. The subset
of computable numbers inside the reals is measure zero, but luckily
K\"ahler moduli arising from moduli stabilization are of this kind,
since moduli stabilization amounts to solving some systems of integer
equations, which can be done numerically to arbitrary
precision. Notice that this idea is general: even if in this paper we
center on diophantine systems, we are not restricting ourselves to
just integer equations, but rather to equations involving computable
numbers, which are in any case those that will appear in satisfactory
top-down approaches to physics.

In the presence of the bundle $\cL$, \eqref{eq:charged-modes-no-L}
gets extended in the following way:
\begin{align}
  (\alpha, \bar\beta) \in (H^0(\cC, \ov \cL\otimes B\otimes
  K_\cC^{1/2}), H^1(\cC, \ov \cL\otimes B\otimes K_\cC^{1/2}))
\end{align}

\subsection{Generalization to F-theory}
\label{sec:F-theory}

The previous discussion seems to admit a relatively straightforward
lift to F-theory, although some of the details are still under active
investigation. For completeness, and since F-theory model building is
an active area of current research (starting with
\cite{Donagi:2008ca,Beasley:2008dc,Beasley:2008kw,Donagi:2008kj}, see
\cite{Weigand:2010wm} for a review of some of the recent
developments), let us sketch how the discussion needs to be modified
in this case. In the context of this paper, where we mostly discuss
complete intersections on toric ambient spaces, there are systematic
methods for uplifting the IIB discussion
\cite{Collinucci:2008zs,Collinucci:2009uh,Blumenhagen:2009up,Cvetic:2009ah,Blumenhagen:2010ja,Cvetic:2010rq}. One
can also gain some understanding when a dual heterotic description is
available \cite{Donagi:2010pd}. In the generic case not coming from a
weakly coupled IIB compactification or the heterotic string, the
counterparts of some important notions in the discussion above are
still incompletely understood, so this review is necessarily
incomplete.

\subsubsection*{Neutral zero modes}

Euclidean D3-brane instantons lift to euclidean M5 branes in F-theory
wrapping a vertical divisor $D$ of a Calabi-Yau 4-fold $X$. Neutral
zero modes on the instanton worldvolume $D$ are in one-to-one
correspondence with representatives of the cohomology groups
$H^i(D,\cO_D)$ \cite{Witten:1996bn}. See
\cite{Blumenhagen:2010ja,Donagi:2010pd} for the detailed map between
the $H^i(D,\cO_D)$ and the IIB modes in
table~\ref{table:O(1)-modes}. The $\theta^\alpha$ modes we discussed
above live in $H^0(D,\cO_D)$, which must be 1 in order for the $M5$
instanton to contribute to the superpotential. One can then construct
a simple index formula that encodes a necessary condition for an
instanton to contribute to the 4d superpotential \cite{Witten:1996bn}:
\begin{align}
  \label{eq:arithmetic-genus}
  \chi(D,\cO_D) = \sum_{i=0}^3 (-1)^i h^i(D,\cO_D) = 1.
\end{align}
As in section~\ref{sec:neutral-modes} this equation admits an
expression as a diophantine equation using the Riemann-Roch
formula. Write $D=\sum d_iD_i$, with $D_i$ a basis of divisors of
$X$. Then~\eqref{eq:arithmetic-genus} ends up being a degree four
equation on the $d_i$ with rational coefficients
\begin{align}
  \chi(D,\cO_D) = m_{ijkl} d_i d_j d_k d_l + n_{ijk} d_i d_j d_k +
  l_{ij} d_i d_j + p_i d_i=1.
\end{align}
which can be made into an equation with integer coefficients by
multiplying both sides by an appropriate integer in order to cancel
denominators.

Notice that we expect the space of F-theory compactification to be
larger than the space of perturbative IIB compactifications, so the
precise details of the problem to solve are different: we are trying
to construct an algorithm to solve a larger set of degree four
equations than in type IIB.

\subsubsection*{Charged zero modes and worldvolume fluxes}

The discussion here is limited by our understanding of worldvolume
dynamics in theories with dyonic vortices. Instantons that lift
straightforwardly from IIB have worldvolume dynamics very similar to
their perturbative IIB description, and can be analyzed along similar
lines to section~\ref{sec:charged-modes}
\cite{Heckman:2008es,Marsano:2008py,Blumenhagen:2010ja,Cvetic:2010rq}.\footnote{Claims
  to the contrary have been recently been made in
  \cite{Donagi:2010pd}. There are cases where one can compare strong
  and weakly coupled descriptions of D-brane instanton effects
  \cite{Forcella:2008au,Aganagic:2007py,GarciaEtxebarria:2008iw,Blumenhagen:2010ja,Cvetic:2010rq},
  and those cases support the agreement between the IIB and F-theory
  descriptions (excepting possible lifting of pairs of zero modes not
  protected by perturbative shift symmetries). F-theory backgrounds
  with no weakly coupled limit may be qualitatively different, and the
  discussion may have to be modified for those.}  Some cases in which
the local geometry includes exceptional singularities can also be
analyzed \cite{Blumenhagen:2010ja}, although it is not clear how to
build a global description in general cases where mutually non-local
zero modes are present (see \cite{Cvetic:2009ah} for a discussion of
some of the subtleties involved).

Similarly, worldvolume fluxes on the F-theory instanton are only well
understood in the case where a IIB limit exists. It is nevertheless
important to remark that there have recently been important advances
in understanding the lift of worldvolume fluxes on D7-branes, see for
example
\cite{Marsano:2009gv,Blumenhagen:2009yv,Grimm:2009yu,Grimm:2010ez,Marsano:2010ix}. Some
of these developments may also be applicable to D-brane instanton
physics.

\medskip

As we have just seen, while the general structure in F-theory seems to
be analogous to the type IIB one, many important details are still
being elucidated, and thus in order to be as explicit as possible we
refrain from discussing F-theory compactifications in what
follows. Nevertheless, once the dust settles and the non-perturbative
effects in F-theory become well understood, we expect the F-theory
discussion to be essentially equivalent in its computability structure
to the IIB discussion. It will thus just strengthen the points we make
below by providing a larger variety of vacua to consider.

\section{General computability}
\label{sec:hilbert}

The previous discussion shows that the ingredients involved in
answering questions about the non-perturbative F-terms are all of a
similar type. In particular, there are important necessary conditions
for superpotential contributions that can be encoded in index
formulas, and we showed above that these give rise to diophantine
equations. One may thus hope that the systematic study of diophantine
equations (which are well studied objects in number theory) may shed
some light on the general problem of computability in the string
theory landscape.

Nevertheless, there are various complications with this
viewpoint. First, it is evident that not all questions reduce
straightforwardly to the study of diophantine equations. One example
we have already mentioned in section~\ref{sec:worldvolume-flux} is the
computation of the supersymmetry conditions on an instanton with
worldvolume flux. More generally, one may ask questions that do not
reduce simply to index formulas, but which do nevertheless make
perfect sense as questions about non-perturbative F-terms. In these
cases, while index conditions may provide very stringent necessary
conditions, the ultimate criterion for computing superpotentials
requires computing equivariant line bundle cohomology, and no closed
formulas exist in general.

It turns out that diophantine equations \emph{do} indeed provide a
unifying framework in which to ask some general questions about
computability, but we need to expand our viewpoint and use some very
non-trivial results in number theory. We proceed to do so in this
section.

\medskip

Let us start by discussing a seemingly cumbersome, but ultimately very
illuminating way of setting up the problem of computing
non-perturbative F-terms in string compactifications. Namely: build
the non-perturbative superpotential of the chosen Calabi-Yau
background by iterating over the set of BPS instantons on that
background, and for each instanton decide which coupling it
contributes to.

Needless to say, this approach has a number of drawbacks. Most
importantly, the procedure takes infinite time, which makes it
unsuitable for any study of the dynamics of the landscape. We could
make the procedure take a finite time if we impose some cut-off on the
wrapping numbers of the instanton in terms of basis
divisors. Physically this approach is easily justified by noticing
that instantons have a suppression factor exponential in the wrapping
number, so beyond some particular wrapping number non-perturbative effects will
not be visible in any finite-precision experiment. This approach by
truncation is easy to implement on a computer, and often gives most of
the information one requires.

Nevertheless, one may not be satisfied by this partial answer by
truncation, and would desire to have a more systematic procedure for
computing non-perturbative F-terms exactly. We are particularly
interested in \emph{whether} such a procedure can exist. It is in the context
of answering this question that the connection to diophantine
equations really shines, since it gives a description of the problem
amenable to analysis using powerful results in number theory.

Before going into the somewhat technical analysis below, let us
summarize the main findings in plain and somewhat imprecise
language. Suppose that, as one would hope, there exists a technique for computing exact
non-perturbative F-terms on arbitrary IIB/F-theory large volume
Calabi-Yau compactifications. We do not require that the procedure is
in any way based on solving the conditions of
section~\ref{sec:systematics} (or more generally any kind of
diophantine equations), only that it is able to answer in finite time
any decision problem involving the
exact F-terms, such as ``\emph{Does every K\"ahler modulus appear at
  least once in the superpotential?}'', ``\emph{Does the
  superpotential include terms involving the K\"ahler moduli $T_1$ and
  $T_2$?}'', or perhaps more exotically ``\emph{Is there any four
  fermion term involving the charged field $X_1$?}''. 
  
Depending on the exact set of questions one is interested in asking,
the final answer for our computability analysis may be different.  We
will therefore be somewhat unspecific in our approach, leaving the set
of questions arbitrary. Given a suggested procedure for computing
non-perturbative F-terms, one has the set of questions that the
procedure can answer, and thus the analysis below will encode whether
the procedure can be extended to the whole set of large volume
IIB/F-theory vacua. In order to be specific, though, we will have in
mind a procedure for computing non-perturbative F-terms such that it
can solve in finite time any decision problem involving
non-perturbative F-terms.\footnote{Decision problems are questions
  having a well defined ``yes'' or ``no'' answer. The class of
  physical questions is generically richer than this, but decision
  problems suffice for our analysis.}

We will argue below that a procedure applicable to the entire
landscape is unlikely to exist, given some genericity conditions on
the space of string compactifications. The argument will be a
refinement and formalization of the following. Consider any such
procedure. As we saw in section~\ref{sec:systematics}, such a
procedure furnishes a way of determining whether a certain system of
integer equations has a solution. Assume the set of equations is
generic enough, such that a procedure for determining whether an
element of the set has a solution exists if and only if there is a
procedure for determining whether an arbitrary system of integer
equations has a solution. It is known, due to the negative solution to
Hilbert's 10th problem (which we review below), that no such procedure
exists for determining solvability of arbitrary integer equations.

\subsection{Formal analysis}

The previous paragraph gives the gist of the argument, but it is
rather imprecise. In this section we will formalize the argument and
give a well defined characterization of the problem of computability
of F-terms. In order to make the conjecture precise we need to
introduce some background material in computability theory and formal
logic. A very readable review with further references can be found in
\cite{MatiyasevichBook}.

\subsection*{Diophantine sets}

Recall that a diophantine equation (over the integers) is a polynomial
equation with integer coefficients such that the unknowns are also
integer numbers. A famous example is:
\begin{equation}
  x^n + y^n = z^n
\end{equation}
with $x,y,z\in\bZ$. In general, we will be dealing with systems of
equations of different orders. A couple of important and elementary
facts about diophantine equations are that every system of high order
equations can be made into a system of quadratic equations by
introducing enough new variables, and that every system of equations
$f_1=\ldots=f_k=0$ is equivalent to the single equation $\sum f_i^2 =
0$. This implies that, without loss of generality, we can discuss the
case of a single diophantine equation of degree four at most.

Let us also introduce the related notion of \emph{diophantine set},
defined as follows: consider a diophantine equation
$f(a_1,\ldots,a_n;x_1,\ldots,x_m)=0$ such that $a_i,x_i\in \bZ$. The
diophantine set $\cF$ associated to $f$ is the set of $a_i$ such that
$f=0$ has a solution. More formally:
\begin{align}
  \cF[f] = \{(a_1,\ldots,a_n) : \exists (x_1,\ldots,x_m) | f(a_1,\ldots,a_n;x_1,\ldots,x_m)=0\}
\end{align}

Instead of thinking about diophantine sets, it will sometimes be
useful to think about properties defining a diophantine set, we refer
to this as \emph{diophantine properties}. As an example, the set
$2\bZ$ of even numbers, or equivalently the property $Even(x)=x\in
2\bZ$, admits the diophantine representation:
\begin{align}
  x\in 2\bZ \iff \exists y|2y = x
\end{align}

\subsection*{Recursive and recursively enumerable sets}

It is also useful to introduce the notion of a \emph{recursively
  enumerable set}. A set $\cR_{\infty}$ is recursively enumerable if
there is a classification algorithm such that for each input $x$ it
halts if and only if $x\in\cR_{\infty}$. An equivalent definition is
that it is possible to construct an algorithm that, possibly in
infinite time, lists all elements of the set. This definition allows
for the classification algorithm to run indefinitely, but in practice,
we need the algorithm to be able to finish in a finite amount of
time. If a classification algorithm finishing in finite time exists,
the set is called \emph{recursive} (or computable).

We have the following result by Matiyasevich (building on earlier work
of Davis, Putnam and Robinson), connecting diophantine and
recursively enumerable sets:
\begin{thm}[Matiyasevich 1970] A set is recursively enumerable $\iff$ it is diophantine.
\end{thm}
In what follows we will thus freely switch between \emph{diophantine}
and \emph{recursively enumerable} when talking about sets. As done
above, we can also introduce the notion of a \emph{recursive
  enumerable property} $\cP$ as a property defining a recursively
enumerable set. If the property $\cP$ holds for an element $x$, we
must be able to determine so in finite time.

We are now in a position to introduce the main two objects of
interest: the diophantine set $\fC_\cP$ of Calabi-Yau spaces defined
by the diophantine property $\cP$, and the \emph{oracle} $\fP$. Let us
start by $\fC_\cP$. We think of $\cP$ as a recursively enumerable
property having to do with the non-perturbative F-terms on the
Calabi-Yau compactification of interest. The questions about K\"ahler
moduli mentioned in the introduction to this section furnish typical
examples of diophantine properties $\cP$. In more generality, any
question boiling down to the existence of a (finite) set of instantons
with the right spectrum of zero modes in a given background is
recursively enumerable,\footnote{Though whether a particular instanton
  contributes to the superpotential or not can be determined in finite
  time, there are an infinite number of possible instantons, so that
  going through each one (in the absence of some better algorithm)
  makes the set recursively enumerable, rather than recursive.} simply
by going through the whole spectrum of instantons and verifying using
the methods reviewed in section~\ref{sec:systematics} whether the
condition is satisfied or not.\footnote{In
  section~\ref{sec:systematics} we discussed indices in generality,
  but discussed the calculation of line bundle cohomology only in the
  case where the geometry is constructible as complete intersections
  in a toric variety. We can generalize the discussion in this section
  to arbitrary Calabi-Yau spaces if we assume that there is a general
  procedure for computing the spectrum of zero modes for any divisor
  that always finishes in finite time.}

Since $\cP$ is by construction a diophantine property of the set of
compactifications, there is an associated diophantine equation that
has solution if and only if the compactification $X$ has the property
$\cP$. That is,
\begin{align}
  \cP(X) \iff \exists x_1,\ldots,x_n|d_{(\cP,X)}(x_1,\ldots,x_n)=0,
\end{align}
where $d_{(\cP,X)}$ is a diophantine equation depending on $\cP$ and
$X$. Equivalently, we can think of $d_{(\cP,X)}$ as encoding the
algorithm that will finish in finite time if and only if $\cP(X)$
holds, which must exist since $\cP$ is recursively enumerable. 
The set $\fC_\cP$ is then defined as the diophantine set of
Calabi-Yau spaces where $\cP$ holds. This set is diophantine as a
subset of the set of string compactifications $\fC$. Notice that
$\fC_\cP$ thus constructed can also be seen as a $\cP$-dependent
subset of the set of diophantine equations with solution, which will be
important below. 

The oracle $\fP$ for the property $\cP$ is defined as an object that
for each Calabi-Yau space $X$ determines whether or not $X\in\fC_\cP$
in finite time.  In this language, the problem of whether an algorithm
exists for computing F-terms exactly can be recast as whether
$\fC_\cP$ is recursive for all $\cP$ relevant to F-terms, or
equivalently whether an algorithmic oracle $\fP$ exists for all $\cP$
relevant to F-terms.  Equivalent questions about the recursivity of
$\fC_\cP$ are interesting for other physical properties $\cP$.

The space $\fC$ of compactifications is somewhat abstract, so let us
formulate the problem (equivalently) in terms of diophantine
equations. Denote by $\fS_{(\cP,X)}$ the set of diophantine equations
having a solution coming from string theory and the property
$\cP$. This set is isomorphic to $\fC_\cP$, but now we are dealing
with solvable diophantine equations $d_{(\cP,X)}$ instead of string
vacua $X$ as elements of the set. This set of solvable $d_{(\cP,X)}$
is a subset of all possible $d_{(\cP,X)}$, which we denote
$D_{(\cP,X)}$ and which is isomorphic to $\fC$.  The question of
interest is whether $\fS_{(\cP,X)}$ is recursive as a subset of
$D_{(\cP,X)}$ for all $\cP$.

\subsection*{Relation to the solution of Hilbert's 10th problem}

As we just discussed, for each property pair $(\cP,X)$ we have a
diophantine equation $d_{(\cP,X)}=0$ that will have a solution if and
only if $\cP(X)$ holds. Consider the set $\fS$ of all diophantine
equations having a solution. Let us denote in addition $D$ as the set
of all diophantine equations. It is easy to see that $\fS$ is
diophantine in $D$: for an element $d\in D$ try all possible values of
the unknowns in some order, if $d\in\fS$ then this process will
eventually halt, and it will not halt if $d\notin\fS$. Our set
$\fS_{(\cP,X)}$ is a subset of $\fS$. Since we have a number of different
inclusions of interest, let us briefly review them. First of all, it
is clear that
\begin{align}
  \fS_{(\cP,X)} \subseteq D_{(\cP,X)}
\end{align}
and that the inclusion is diophantine. Similarly,
\begin{align}
  \fS_{(\cP,X)} \subseteq \fS
\end{align}
but notice that now $\fS_{(\cP,X)}$ is not clearly diophantine as a
subset of $\fS$, since we have not specified a way of determining
whether a given solvable diophantine equation comes from string
theory. If the set of Calabi-Yau spaces is recursive, it follows
easily that $\fS_{(\cP,X)}\subseteq\fS$ is also recursive. Another way
of saying this is that we are assuming that there is in principle a
process that can determine in finite time whether a given diophantine
equation with solution comes from a pair $(\cP,X)$ as constructed
above, for fixed $\cP$. It is often conjectured that there is a finite
number of families of Calabi-Yau threefolds \cite{Yau-Survey}, which
we take as an assumption. In this case $\fS_{(\cP,X)}$ would be
trivially recursive in $\fS$, simply by enumeration of the set of
string vacua. Similarly, even if the space of all possible
compactifications $\fC$ turns out to be infinite,
$\fS_{(\cP,X)}\subseteq \fS$ is still recursive as long as $\fC$
decomposes into a finite sum of parametric families, since the
parametric family to which a diophantine equation belongs can be
determined in finite time.

Let us thus assume for the sake of the argument that the space of
string theory vacua is recursive. Given a diophantine equation $d\in
D$ and a condition $\cP$, this also allows one to determine in finite
time whether or not $d$ comes from string theory, that is, whether it
is one of the diophantine equations $d_{(\cP,X)}$. It renders the
inclusion
\begin{align}
  D_{(\cP,X)} \subseteq D
\end{align}
recursive and
\begin{align}
  \fS_{(\cP,X)} \subset D.
\end{align}
recursively enumerable.

It is easy to see that if a recursively enumerable set $\cS$ is
recursive then its complement $\overline\cS$ is recursively
enumerable. The converse is also true: if the complement
$\overline\cS$ of a diophantine set $\cS$ is diophantine, then $\cS$
is recursive. To see this, imagine running the algorithms for
determining whether an element $x$ is in $\cS$ or $\overline\cS$ in
parallel, one instruction at a time. Since both sets are diophantine,
the algorithm will eventually stop and give the correct answer for
whether $x\in\cS$ or $x\notin\cS$.

We are now in a position to make our intuitive discussion about the
relation with Hilbert's 10th problem precise. A simple argument
\cite{MatiyasevichBook} shows that $\overline\fS$ is not diophantine
in $D$, and thus $\fS$ is not recursive. This is the (negative) solution
to Hilbert's tenth problem, since it is equivalent to the statement that
there is no algorithm which can determine in finite time whether or not
an arbitrary diophantine equation has a solution. In our context, we are
dealing with $\cP$-dependant subsets of $\fS$, or equivalently
\begin{align}
\label{eq:complement-inclusion}
\overline{\fS}\subseteq\overline{\fS_{(\cP,X)}}.
\end{align}
The intuitive notion of genericity of the space of Calabi-Yau vacua,
and the possible properties $\cP$ to ask for, then amounts to the
statement that $\overline{\fS_{(\cP,X)}}$ is non-diophantine for some
$\cP$ as a subset of $D$.

Notice that from this viewpoint, this genericity condition seems
rather natural, as in scanning over the space of properties $\cP$ and
the landscape we are scanning over large subsets of the space of
possible diophantine equations. A recursive procedure for determining
whether $\cP$ is true for an arbitrary Calabi-Yau is now encoded as a
procedure for determining whether some rather generic diophantine
equations have a solution. This leads us to conjecture the following:
\begin{conjecture}[Non-computability]
  $\fS_{(\cP,X)}$ is non-recursive as a subset of $D$ for some $\cP$.
\end{conjecture}

\subsection{Discussion}
\label{sec:hilbert-discussion}

An important assumption above was that the set of relevant string
compactifications is somehow enumerable. This is often implicitly or
explicitly taken to be true, and in fact there are various reasons to
think that the relevant space of geometric compactifications is finite
\cite{Yau-Survey}. In string theory, in addition to the background
geometry, there are also typically other elements involved in defining
the vacuum, such as branes, orientifolds or fluxes. It is also
believed that there are large but finite bounds on the number of flux
vacua of string theory (see \cite{Douglas:2006es} for a review).

Let us briefly consider the case that an enumeration procedure for
string vacua is not constructible, even in principle. This would make
the formalization above not valid since we cannot conclude anymore
that $\fS_{(\cP,X)}\subset D$ is diophantine. Nevertheless, at least
intuitively, it is hard to imagine an algorithmic oracle $\fP$
existing for answering any question about the structure of elements of
a non-diophantine set. So if string vacua were not enumerable in
principle this would seem to rather strengthen the case for
non-computability.

\medskip

Notice that the non-computability conjecture came most naturally as a
statement about $\fS_{(\cP,X)}$ as a subset of $D$, but physically we
are rather interested on the properties of $\fS_{(\cP,X)}$ as a subset
of $D_{(\cP,X)}$ instead. It turns out that the non-computability
conjecture above implies the non-recursiveness of $\fS_{(\cP,X)}$ in
$D_{(\cP,X)}$: assume that $\fS_{(\cP,X)}$ is recursive as a subset of
$D_{(\cP,X)}$. Then, given a diophantine equation $d\in D$ we can
determine (by assumption of recursivity of $D_{(\cP,X)}\subseteq D$)
in finite time whether it is in $D_{(\cP,X)}$ or not. If not, it is
obviously not in $\fS_{(\cP,X)}$ either. If it is in $D_{(\cP,X)}$, by
assumption of recursivity of $\fS_{(\cP,X)}$ in $D_{(\cP,X)}$, we can
determine in finite time whether it is solvable or not. Thus,
recursivity of $\fS_{(\cP,X)}$ in $D_{(\cP,X)}$ implies recursivity of
$\fS_{(\cP,X)}$ in $D$, and thus runs contrary to the non-computability
conjecture above. Therefore, if the conjecture holds then it is also
true that $\fS_{(\cP,X)}$ is non-recursive in $D_{(\cP,X)}$.

\medskip

Finally, we would like to emphasize once more that it is the
\emph{computability structure} of the problem that connects the
problem of computing F-terms to the abstract discussion of diophantine
sets, and not just the fact that some simple problems involving index
formulas admit a simple diophantine representation, although this was
our original motivation. Any question that one can ask, and whose
solution can be encoded in terms of a mechanical procedure on the
space of instantons, admits a formulation in the framework of this
section. In particular, as we have shown explicitly, any question that
can be formulated uniquely in terms of the structure of zero modes of
various instantons is part of the set $\cP$.

\section{Exact superpotentials for special manifolds}
\label{sec:elliptic-fibrations}

In the previous section we took a look at physical properties $\cP$
across the string theory landscape. One interesting class of properties
$\cP$ involves questions related to the structure of F-terms. 
We saw that the generic problem admits a
neat formulation in terms of computability theory, and this hints to
some general properties of the landscape. An important part of the
link is the realization that diophantine equations provide a universal
description of the computational structure involved.

In this section we want to look to this result from a complementary
angle: instead of assuming generic backgrounds and studying the class
of equations that arise, we will impose a particular structure for the
equations, such that they are solvable, and ask which parts of the
landscape give rise to such kind of equations for a particular
property $\cP$.

Such a structure would determine \emph{solvable} $\cN=1$ backgrounds,
in the sense that one can explicitly solve for all instantons
satisfying a given condition. Due to its physical interest, the
condition that we aim to solve for is that a given instanton
contributes to the uncharged superpotential. We do not give the
general criterion necessary for solvability, but give instead an
interesting family of solvable backgrounds. The basic property common
to this class of solvable manifolds turns out to be that their
intersection form on the Calabi-Yau factorizes, by which we mean that
there is a divisor which participates in every non-zero triple
intersection on the Calabi-Yau. Such intersection forms are commonly
seen for elliptically fibered threefolds in Weierstrass form, though
it also occurs for manifolds which do not admit an elliptic
fibration. We give examples of both situations, as well as an example
of an elliptic threefold which does \emph{not} factorize and a
factorizing threefold which is not an elliptic fibration.

\medskip

At this point we wish to explicitly state the conditions which
guarantee the solvability of all uncharged instanton corrections
to the superpotential:
\begin{solvability}{(Uncharged superpotential corrections)}\\
  Given the set $\cS$ of smooth Calabi-Yau compactifications with
  intersection three-form of type $I_X=Df_2$ for some divisor $D$ and
  $f_2$ a quadratic polynomial of divisors, and $D_{D7}$ having no
  component along $D$, then the set $\cS_\cP\subset \cS$ for which any
  given property $\cP$ of the uncharged superpotential holds is
  recursive.
\end{solvability}
We remind the reader that the properties $\cP$ that we are interested
in are those that can be formulated in terms of the zero modes of the
instantons on each vacuum. The argument for why these conditions allow
for a solvability algorithm is roughly as follows: an instanton which
generates the uncharged superpotential contribution of interest has no
charged zero modes, which requires that it not intersect the $D7$
branes or that it do so at a $\bP^1$.\footnote{Recall from
  section~\ref{sec:charged-modes} that charged zero modes are counted
  by the cohomology groups $H^0(\cC,\cL)$ and $H^1(\cC,\cL)$ for some
  bundle $\cL$ that depends on worldvolume fluxes and
  $K_\cC^{\frac{1}{2}}$.  Generically such zero modes exist, except in
  the special case where $\cC=\bP^1$ and $\cL=\cO(-1)$, as in that
  case both cohomologies vanish. This is not so exotic, as $D7$ branes
  often intersect a divisor at a $\bP^1$, and in the absence of
  worldvolume flux $\cL=K_\cC^{\frac{1}{2}}=\cO(-1)$.}  If the
solvability conditions are satisfied, the corresponding diophantine
equations factorize into a product of two linear diophantine
equations, rendering it possible to solve for all divisors which do
not intersect the $D7$ branes or do so at a $\bP^1$. We make this
explicit in non-trivial examples, given by Calabi-Yau elliptic
fibrations over $\bP^2$, $dP_2$ and $\bF_n$.

\medskip

In fact, for Calabi-Yau elliptic fibrations there exist a set of
conditions which ensure that the intersection form factorizes. For
generic manifolds, studying these conditions does not necessarily
offer any advantage over studying the intersection form directly, but
in the case of elliptic fibrations realized as a hypersurface in a
toric variety, the conditions make factorization rather transparent,
and also suggests a method for constructing an elliptic Calabi-Yau in
Weierstrass form over a given base $\cB$.\footnote{Throughout this
  paper all bases $\cB$ are toric. It is also possible to construct
  factorizing manifolds for cases where $\cB$ is a hypersurface in a
  toric variety, and thus the elliptic threefold is a complete
  intersection Calabi-Yau. In such a way one can construct elliptic
  threefolds with base $dP_5$, $dP_6$, $dP_7$, and $dP_8$ which
  exhibit a factorizing intersection form.}

Consider a Calabi-Yau elliptic fibration (with section) $\pi: X\to
\cB$. We call the Stanley-Reisner ideal of the base and threefold
$SRI_\cB$ and $SRI_X$ respectively, and we denote divisors in the base
as $C_i$, as they are curves in the threefold.  Now suppose that the
structure of the fibration is such that it satisfies three basic
assumptions:
\begin{itemize}
\item $X$ satisfies $h^{1,1}(X) = h^{1,1}(\cB) +1$,  which is equivalent to saying that
$Div(X)$ has one more generator than $Div(\cB)$. 
\item Every generator of $SRI_\cB$ is also a generator of $SRI_X$.
\item Linear equivalence of base divisors is preserved in $X$. That is, if $C_i\sim C_j$,
then $\pi^{-1}(C_i)\sim\pi^{-1}(C_j)$.
\end{itemize}
These conditions are satisfied in many concrete examples. For example,
the first condition is always satisfied when $X$ is a hypersurface in
a toric variety $\cA$ with one more $\bC^*$ action than the toric
variety of the base. Notable examples include fibrations where the
fiber is a hypersurface in $\bP^2$ or $\bP_{231}$, which include many
Weierstrass fibrations. The second condition can be checked explicitly
in each topological phase of the threefold, and the third condition is
satisfied by many examples of the kind discussed in this work (see
tables~\ref{table:T2overP2}, \ref{table:T2overFn} and
\ref{table:T2overdP2}) since the homogeneous coordinates corresponding
to $\cB$ are not charged under the additional $\bC^*$ action of $\cA$.
If these conditions are satisfied, then cubic terms in the
intersection ring $\cI_X$ which involve only pullbacks of divisors of
the base must vanish. Thus, any non-vanishing triple intersection of
divisors necessarily involves the divisor $K$ corresponding to the
additional K\"ahler modulus of $X$, and this divisor therefore
factorizes out of the intersection three-form.

\subsection{Factorization exemplified: an elliptic fibration over $\bP^2$}
\label{sec:T2overP2}

We would now like to show the use of factorization in solving for
non-perturbative effects in an example which is non-trivial,\footnote{More trivial examples
would include smooth Calabi-Yau threefolds with a single K\"ahler modulus,
in which the holomorphic genus is a cubic equation in a single variable,
and thus the finite set of instantons satisfying $\chi(D,\cO_D)=1$ can
be algorithmically determined. 
For example, on the quintic threefold $\chi(D, \cO_D) =\frac{5}{12}\,n\,(2n^2+10)$,
where $D=nH$, and therefore no instanton contributes to the superpotential in the absence
of flux.}
yet does not involve tedious calculations which obscure
the point. To this end, let us consider an elliptically fibered
Calabi-Yau threefold over $\bP^2$. The GLSM data of the toric ambient
space $\cA$ in which the threefold is a hypersurface is given in table
\ref{table:T2overP2}. This data can be thought of as being constructed
by taking a $\bP^2$ with homogeneous coordinates $s$, $t$, and $u$
augmented by a $\bP_{231}$ with coordinates $x$, $y$, and $z$. The
Calabi-Yau curve in $\bP_{231}$ gives in this way the elliptic fiber
in a Weierstrass fibration over $\bP^2$. The GLSM charges of $x$, $y$,
and $z$ under $Q_1$ are determined by the Calabi-Yau conditions
associated with the Weierstrass equation.

\TABLE{
\begin{tabular}{c|c|c|c|c|c|c}
	& $s$ & $t$ & $u$ & $x$ & $y$ & $z$\\ \hline
    	$Q_1$ & 1 & 1 & 1 & 6 & 9 & 0 \\
      	$Q_2$ & 0 & 0 & 0 & 2 & 3 & 1 
\end{tabular}
\label{table:T2overP2}
\caption{GLSM charges for a toric ambient space $\cA$ whose Calabi-Yau
  hypersurface is an elliptic fibration over $\bP^2$.}
}

There are two triangulations corresponding to these GLSM charges, one of
which has a Stanley-Reisner ideal given by $SRI = \langle
stu,xyz\rangle$, which is the one considered here. The generators of the $SRI$
in this topological phase are thus simply the generators of $SRI_{\bP^2}$
and $SRI_{\bP_{231}}$.  We define the generators of $Div(\cA)$ to be
$H\equiv D_s$ and $K\equiv D_z$.\footnote{We denote by $D_{x_i}$ the
  divisor given by the coordinate $x_i$ vanishing. For example, $D_s$
  denotes the locus $\{s=0\}$.} In this basis, the Calabi-Yau hypersurface
in $\cA$ has divisor class $18H+6K$ and the intersection form on the
Calabi-Yau is
\begin{equation}
I_X = K(H^2-3HK+9K^2).
\end{equation}
This data is already enough to compute interesting topological
indices, such as the holomorphic genus or the Euler character of a
divisor or curve. Doing so requires calculating the appropriate Chern
classes from the given information. The ones that we will need are
calculated by adjunction to be
\begin{align}
  c(T_X) = 1 + 102H^2 + 69HK + 11K^2 -1628H^3 -1629H^2K - 543HK^2-60K^3
\end{align}
and
\begin{align}
  c(T_D) = 1 - nH -mK + (n^2+102)H^2 + (2nm+69)HK + (m^2+11)K^2,
\end{align}
where $X$ is the Calabi-Yau and $D=nH+mK$ is a divisor in it. The
intersection form on $D$ is given by:
\begin{align}
  I_D = mH^2 + (n-3m)HK + (9m-3n)K^2.
\end{align}

With this information the holomorphic genus is then easily calculated
to be:
\begin{equation}
  \label{eq:P2-fibration-genus}
  \chi(D,\cO_D) = \frac{1}{12} \int_D c_1^2(T_D) + c_2(T_D) =
  \frac{1}{2}(3m^3 - 3m^2n + mn^2 - m + 6n),
\end{equation}
yielding a diophantine equation which is difficult to solve explicitly for the
necessary constraint $\chi(D,\cO_D)=1$. If this was the only
constraint which must be satisfied for an instanton to contribute to
the superpotential, the task of identifying all such instantons would
be difficult indeed. Luckily, in this case some of the other necessary
conditions become simple enough to allow us to solve the problem
exactly.

Introduce a holomorphic orientifold involution $\sigma:s\mapsto
-s$. The only divisor in $\cA$ which is pointwise fixed under $\sigma$
is $D_s$. This determines the location of an $O7$ plane, so that
$[D_{O7}] = H$.  Such an object is magnetically charged under the
Ramond-Ramond zero form, and this charge must be cancelled in the
internal space by the introduction of $D7$ branes. This tadpole
cancellation condition can be expressed in homology as
\begin{equation}
\sum_a N_a([D_a] + [D^{'}_a]) = 8[D_{O7}],
\end{equation}
where $[D_a]$ is the homology class of a divisor wrapped by $N_a$ $D7$
branes and $[D^{'}_a]$ is its image under $\sigma$. In this example it
should be noted that $\sigma$ acts trivially on homology, and
therefore $[D_a]=[D^{'}_a]$, though not necessarily pointwise. Given
that $[D_{O7}]=H$ in this example, one solution to the tadpole
cancellation conditions is to have three $D7$ branes wrapping $D_s$
and one $D7$ brane wrapping $D_t$. If our concern here was with
model-building, this would give rise to an $SO(6) \times Sp(2)$ gauge
group, which could be broken to $U(3) \times U(1)$ by turning on
fluxes.  For our purposes, it suffices to know which divisors the $D7$
branes wrap.

We now would like to classify instantons in this example which do not
have charged zero modes. This occurs when the divisor which an
instanton wraps does not intersect any $D7$ brane or does so at a
$\bP^1$. Generically, an instanton wrapped on $D$ does not intersect
any gauge $D7$ brane if $K_i\cdot D\cdot D_{D7}=0$ for all linearly
independent generators $K_i$ of $Div(X)$ and all $D7$ branes. Since
the divisors which the $D7$ branes wrap in this example are both $H$
homologically, the non-intersection conditions are
\begin{align}
  \begin{split}
    H\cdot D\cdot H &= \, nH^3 + mH^2K = \, m = 0 \\
    K\cdot D\cdot H &= \, nH^2K + mHK^2 = \, n-3m = 0,
  \end{split}
\end{align}
which show that all non-trivial divisors intersect a $D7$ brane.

We now investigate which divisors $D$ intersect a $D7$ brane at a
$\bP^1$. Since $\bP^1$ is the Riemann surface of genus zero, a simple
way to do this is to calculate $\chi(C) = \int_C c_1(T_C) = 2-2g$ for
$C=D\cdot D_{D7}$. For a curve which is a complete intersection in a
Calabi-Yau threefold this simplifies to
\begin{equation}
\label{eqn:euler of curve cy}
\chi(C) = - D\cdot D_{D7} \cdot (D + D_{D7}), 
\end{equation}
and the constraint for intersecting at
$\bP^1$ in this example is therefore
\begin{align}
\label{eqn:euler of curve T2overP2}
\chi(D\cdot H) = m(3m-2n-1)=2.
\end{align}
This is a quadratic inhomogeneous equation in two variables. Generic
inhomogeneous equations are hard to solve, but luckily the left hand
side of~\eqref{eqn:euler of curve T2overP2} factorizes. Since we are
solving over the integers, this simplifies the problem
considerably: 2 is a prime number, and thus either
\begin{align}
  \begin{split}
    m &= \pm 1\\
    3m - 2n - 1 & = \pm 2
  \end{split}
\end{align}
with the signs on both equations agreeing, or
\begin{align}
  \begin{split}
    m &= \pm 2\\
    3m - 2n - 1 &= \pm 1.
  \end{split}
\end{align}
Solving these linear equations is straightforward, the possible
solutions are $(m,n)=(1,0)$, $(-1,-1)$,$(2,2)$, and $(-2,-3)$. As it
is easily verified by substitution in~\eqref{eq:P2-fibration-genus},
the only divisor in this set with $\chi(D,\cO_D)=1$ is $(m,n)=(1,0)$.
Or, in other words, $D=K$. This is the divisor corresponding to the
$\bP^2$ base of the fibration, and thus we also know that
$h^i(K,\cO_K)=0$ for $i>0$. An analysis of equivariant cohomology (see details
in later examples) shows that $h^0_+(K,\cO_K)=1$ and $h^0_-(K,\cO_K)=0$, from which
we conclude that this divisor does indeed contribute to the uncharged superpotential.

\medskip

There is a valuable general lesson to be learned from this
exercise. Equation \eqref{eqn:euler of curve T2overP2} was solvable
precisely because it factored into the product of two linear terms. By
considering \eqref{eqn:euler of curve cy}, it is straightforward to
see that such factorization is guaranteed to occur if some divisor $K$
factorizes out of the intersection three-form and $D_{D7}$ has no
component along $K$. The latter condition is rather common for
Calabi-Yau orientifold compactifications in type IIB with an
intersection form that factorizes.

For Calabi-Yau manifolds of this type, this suggests a generic
prescription for investigating instanton effects systematically. Given
the genus $g$ of $C=D\cdot D_{D7}$, solve the product of linear
equations given by $\chi(C)$ for all divisors $D$ which intersect the
$D7$ brane at a Riemann surface of that genus. Then, check to see if
those divisors satisfy other necessary constraints for superpotential
contribution, such as $\chi(D,\cO_D)=1$. One might be concerned that
this is just replacing one problem with another, since the genus $g$
of $C$ could be any non-negative integer. However, such higher genus
curves are accompanied by large numbers of charged zero modes, which
would give superpotential corrections of high mass dimension. Since
such corrections are irrelevant for low energy physics, it makes sense
to set an upper bound $k$ on the genus of the curve and solve
$\chi(C)=2-2g$ only for $g\le k$.

\subsection{Further examples: $\bF_n$ and $dP_2$ base, a non-factorizing elliptic
threefold, and a non-elliptic factorizing threefold}
\label{sec:further examples}

In this section we present more examples to further illustrate the utility of
factorization in solving for instanton corrections. We consider elliptic threefolds
with $\bF_n$ and $dP_2$ base, where the toric data of the base is augmented by a
copy of $\bP_{231}$, along with the Calabi-Yau conditions coming from the Weierstrass equation.
We also give an example of an elliptic threefold which is a hypersurface in $dP_1 \times \bF_0$,
where the intersection threeform does not factorize, as can be seen intuitively from the
toric data. Finally, we briefly present a well-known example of a factorizing threefold
which admits a $K3$ fibration, but not an elliptic fibration.

\subsubsection*{Elliptic fibrations over $\bF_n$}

Let us start by discussing elliptic fibrations over
$\bF_n$.\footnote{We note that this must be done with some care, as
  the total space of the fibration has singularities for $n>2$, which
  would be interpreted in F-theory as non-abelian gauge symmetry.} The relevant
GLSM data is given in table~\ref{table:T2overFn}. We take the
triangulation corresponding to $SRI=\langle uv,st,xyz\rangle$. A
hypersurface in the toric variety given by this GLSM data satisfies
the conditions for factorization, and indeed we easily find a
factorizing intersection form in the Calabi-Yau hypersurface:
\begin{align}
  I_X =P(MO-2MP-nO^2+(n-2)OP+8P^2).
\end{align}
We take the holomorphic involution to be $\sigma:v\mapsto -v$ which leaves
$D_v$ and $D_u$ pointwise fixed, so that
$[D_{O7}] = [D_u] + [D_v] = nM + 2O$.

\TABLE{
\begin{tabular}{c|cccc|ccc}
	& $s$ & $t$ & $u$ & $v$ & $x$ & $y$ & $z$\\ \hline
   	$M$ & $1$ & $1$ & $n$ & $0$ & $2(n+2)$ & $3(n+2)$ & $0$ \\
    $O$ & $0$ & $0$ & $1$ & $1$ & $4$ & $6$ & $0$ \\ \hline
	$P$ & $0$ & $0$ & $0$ & $0$ & $2$ & $3$ & $1$
\end{tabular}
\label{table:T2overFn}
\caption{GLSM charges for a 4D toric ambient space $\cA$ whose Calabi-Yau hypersurface is an elliptic fibration over $\bF_n$.}
} 

As one would expect, the branes which must be added to cancel
tadpoles are dependent on which Hirzebruch surface is the base. If we
consider wrapping $n_s$ $D7$ branes on $D_s$, $n_u$ $D7$ branes on
$D_u$ and $n_v$ $D7$ branes on $D_v$, tadpole cancellation and the
wrapping of the $D7$ branes on effective divisors requires $n_u + n_v
= 8$ and $n_s+nn_u=4n$ with $n_u \le 4$.  We choose to have $n_u = 3$
branes wrapping $D_u$, $n_v=5$ branes wrapping $D_v$, and $n$ branes
wrapping $D_s$. This gives rise to an $SO(10)\times SO(6) \times
Sp(2n)$ gauge symmetry on the D7 branes.

Writing an arbitrary divisor as $D=mM+oO+pP$, the holomorphic genus is
given by:
\begin{align}
  \label{eq:Fn-genus}
  \begin{split}
\chi(D,\cO_D) &= -\frac{1}{6}(3no^2p - 3nop^2 - 6mop + 6mp^2 + 6op^2\\
&  \phantom{=-\frac{1}{6}(}- 8p^3 + 6no - 12m - 12o + 2p).
\end{split}
\end{align}
Since it again seems rather difficult to solve for all solutions to
$\chi(D,\cO_D)=1$, we impose the additional constraints coming from
requiring the absence of charged zero modes. The conditions for a
divisor $D$ to not intersect the $D7$ branes wrapped on $D_s$ are
given by
\begin{align}
  \begin{split}
  M\cdot D_s \cdot D &= 0\\
  O\cdot D_s\cdot D & = p = 0\\
  P\cdot D_s\cdot D & = o-2p = 0.
  \end{split}
\end{align}
Similar calculations for the intersection of $D$ with the $D7$ branes
on $D_u$ and $D_v$ give the results
\begin{align}
\label{eqn:Fn no intersection}
D_s \cdot D = 0 \qquad &\leftrightarrow \qquad p=0\qquad o=0 \notag \\
D_u \cdot D = 0 \qquad &\leftrightarrow \qquad p=0\qquad m=0 \\
D_v \cdot D = 0 \qquad &\leftrightarrow \qquad p=0\qquad no=m. \notag
\end{align}
The conditions for $D$ to intersect the $D7$ branes wrapping $D_s$,
$D_u$, or $D_v$ at a $\bP^1$ are respectively given by
\begin{align}
\label{eqn:Fn P1 intersection}
\chi(D\cdot D_s) \,&=\, -2p(o - p) \,=\, 2 \notag \\
\chi(D\cdot D_u) \,&=\, p(np - n - 2m + 2p) \,=\, 2\\
\chi(D\cdot D_v) \,&=\, p(2no - np + n - 2m + 2p) \, = \, 2. \notag
\end{align}
Looking to these systems of equations, we again see that these
conditions show that $D$ must not intersect any $D7$ branes or must
intersect every one at a $\bP^1$. Divisors satisfying~\eqref{eqn:Fn no
  intersection} are of the form $D=nN$, with $n$ arbitrary. By
substitution in~\eqref{eq:Fn-genus} we easily see that for such
divisors $\chi(D,\cO_D)=0$, and thus they do not contribute to the
superpotential. On the other hand, $D=P$ satisfies \ref{eqn:Fn P1
  intersection}, as does $D=-P$ for the special case of $n=0$, that
is, when the elliptic fibration has base $\bF_0=\bP^1\times\bP^1$. The
latter has $\chi(D,\cO_D)=-1$, and therefore does not contribute to
the superpotential. For $D=P$, though, $\chi(D,\cO_D)=1$.

We have determined that the only instanton which might have the
correct uncharged zero mode structure to give rise to a neutral
superpotential correction is $D=P$. We now apply the techniques of
appendix \ref{sec:equivariant-cohomology} to directly calculate the
equivariant cohomology $h^i_\pm (D,\cO_D)$. The relevant Koszul
sequences are:
\begin{align}
\label{eqn:T2overFn Koszul}
\begin{split}
  0 \rightarrow \cO_X(-D) \rightarrow \cO_X \rightarrow \cO_D \rightarrow 0 \\
  0 \rightarrow \cO_\cA(-X) \rightarrow \cO_\cA \rightarrow \cO_X \rightarrow 0 \\
  0 \rightarrow \cO_\cA(-X-D) \rightarrow \cO_\cA(-D) \rightarrow
  \cO_X(-D) \rightarrow 0,
\end{split}
\end{align}
and thus we must calculate the corresponding ambient space
cohomologies. The result is that the only non-zero cohomologies are
generated by sections of the form
\begin{align}
&H^0(\cA,\cO_\cA)\sim \text{const} \notag \\ 
H^4(\cA,\cO_\cA(-X)) \sim \frac{1}{stuvxyz}& \qquad H^4(\cA, \cO_\cA(-X-D)) \sim \frac{1}{stuvxyz^2}.
\end{align}
Each section contributes once to the corresponding cohomology, so the
dimension of each cohomology is one. Utilizing the long exact
sequences in cohomology corresponding to the Koszul
sequences~\eqref{eqn:T2overFn Koszul} gives
$h^i(D,\cO_D)=(1,0,0)$, in agreement with the holomorphic genus
calculation.

Finally, we must calculate how the cohomology $h^0(D,\cO_D)=1$ splits
into equivariant cohomology. Again using the techniques discussed in
appendix \ref{sec:equivariant-cohomology}, the $\bZ_2$ action on the
geometry given by $\sigma:v\mapsto -v$ induces an action on the
sections, from which is can be seen that the relevant group character
is $\chi_g(H^0(\cA,\cO_\cA)) = (1,1)$.  This gives that
\begin{equation}
\chi_g(H^0(D,\cO_D)) = (1,1),
\end{equation}
and thus $h^0_+(D,\cO_D)=1$ and $h^0_-(D,\cO_D)=0$. This indicates
that the $\ov\tau$ mode has been projected out, as expected since $P$
is an orientifold invariant divisor. From these equivariant cohomology
calculations, we see that $P$ satisfies the necessary and sufficient
conditions on uncharged zero modes for an instanton wrapping it to
give an uncharged superpotential contribution.

\subsubsection*{Elliptic fibration over $dP_2$}

We now turn to another example of a concrete calculation in a
non-trivial case. The GLSM data for this elliptic fibration over
$dP_2$ is given in table~\ref{table:T2overdP2}. It is straightforward
to see the $dP_2$ and $\bP_{231}$ structure from the GLSM data, which
we have partitioned into the relevant quadrants. (This manifold was
constructed in \cite{Blumenhagen:2008zz} as a del Pezzo transition of
$\bP_{11169}[18]$.) The intersection form on the Calabi-Yau
hypersurface $X$ is given by
\begin{equation}
  \label{eq:dP2-fibration-IX}
  I_X = P(MN+MO-MP-NP-OP-M^2-N^2-O^2+7P^2)
\end{equation}
Finally, the Stanley-Reisner ideal is given by $SRI = \langle x_1x_3,x_1x_4,x_2x_3,x_2x_5,x_4x_5,x_6x_7x_8\rangle$.

Writing an arbitrary divisor as $D = mM + nN + oO + pP$ and calculating
$c(T_D)$ by adjunction as in section \ref{sec:T2overP2}, the
holomorphic genus is calculated to be
\begin{align}
  \begin{split}
\chi(D,\cO_D) &= -\frac{1}{6}(3m^2p - 6mnp + 3n^2p - 6mop + 3o^2p + 3mp^2\\
&  \phantom{=-\frac{1}{6}(}+3np^2 + 3op^2 - 7p^3 - 6m - 6n - 6o + p).
\end{split}
\end{align}

\TABLE{
\begin{tabular}{c|ccccc|ccc}
	& $x_1$ &$x_2$ &$x_3$ &$x_4$ &$x_5$ &$x_6$ &$x_7$ &$x_8$ \\  \hline
    $M$ & $1$ & $1$ & $1$ & $0$ & $0$ & $6$ & $9$ & $0$ \\
    $N$ & $0$ & $1$ & $0$ & $0$ & $1$ & $4$ & $6$ & $0$ \\
	$O$ & $1$ & $0$ & $0$ & $1$ & $0$ & $4$ & $6$ & $0$ \\ \hline
	$P$ & $0$ & $0$ & $0$ & $0$ & $0$ & $2$ & $3$ & $1$
\end{tabular}
\label{table:T2overdP2}
\caption{GLSM charges for a 4D toric ambient space $\cA$ whose Calabi-Yau hypersurface is an elliptic fibration over $dP_2$.}
} We take the orientifold involution to be $\sigma: x_2 \mapsto
-x_2$. The divisors which are pointwise fixed by this involution are
$D_2$ and $D_5$ and are thus are wrapped by $O7$ planes, giving
$[D_{O7}] = [D_2]+[D_5] = M+2N$. The Ramond-Ramond tadpole can be
cancelled by introducing five $D7$ branes on $D_5$, three on $D_2$ and
one on $D_3$, giving rise to gauge group $SO(10)\times SO(6) \times
Sp(2)$ before turning on fluxes, which can break the first factor to
obtain a $SU(5)$ GUT. We wish to examine under what conditions $D$
intersects one of the $D7$ branes. We again use the techniques
described in section \ref{sec:T2overP2}, obtaining the conditions for
$D$ to intersect the $D7$ brane stack on $D_2$:
\begin{align}
  \begin{split}
  M\cdot D_2 \cdot D &= 0\\
  N\cdot D_2 \cdot D &= 0\\
  O\cdot D_2\cdot D & = p = 0\\
  P\cdot D_2\cdot D & = o-2p = 0.
  \end{split}
\end{align}
Similar calculations for the intersection of $D$ with the $D7$ branes
on $D_3$ and $D_5$ give the results
\begin{align}
\label{eqn:no intersecton}
D_2 \cdot D = 0 \qquad &\leftrightarrow \qquad p=0\qquad o=2p \notag \\
D_3 \cdot D = 0 \qquad &\leftrightarrow \qquad p=0\qquad m=n+o \\
D_5 \cdot D = 0 \qquad &\leftrightarrow \qquad p=0\qquad m=n. \notag
\end{align}

As before, charged modes can also be absent when $D$ intersects a $D7$
brane at a $\bP^1$. Calculating the Euler character for the curves
where $D$ intersects each of the divisors which the $D7$ branes wrap
gives the conditions
\begin{align}
\label{eqn:P1 intersection}
\chi(D\cdot D_2) \,&=\, -2p(o - p) \,=\, 2 \notag \\
\chi(D\cdot D_3) \,&=\, p(2m - 2n - 2o + p + 1) \,=\, 2\\
\chi(D\cdot D_5) \,&=\, -p(2m - 2n - p - 1) \, = \, 2 \notag
\end{align}
for intersection at a $\bP^1$. The unique solution for intersecting all three
at a $\bP^1$ is $m=n$, $o=0$ and $p=1$.

Now, the condition for the absence of charged modes is that an
instanton wrapped on $D$ must either not intersect each $D7$ brane or
must do so at a $\bP^1$. Satisfying equation \eqref{eqn:P1
  intersection} requires $p\ne 0$, though, which makes it impossible
to satisfy equation \eqref{eqn:no intersecton}. Thus, the divisors $D$
which have no charged modes either intersect none of the $D7$ branes
or intersect every $D7$ brane at a $\bP^1$. The absence of charged
modes restricts $D$ to be of the form
\begin{align}
  D &= mM + mN+P,
\end{align}
or alternatively
\begin{align}
  D &= mM + mN,
\end{align}
where divisors of the first form intersect the $D7$ branes and
divisors of the second form do not. The holomorphic genus of these
divisors takes the dramatically simplified form
\begin{align}
  \begin{split}
    \chi(D,\cO_D) &= m+1 = 1 \\
    \chi(D,\cO_D) &= 2m = 1
  \end{split}
\end{align}
Thus, the simple conclusion of these results is that there is no
instanton which does not intersect the gauge branes that contributes
to the superpotential, and that
\begin{equation}
D = P = D_8
\end{equation}
intersects all $D7$ branes at a $\bP_1$ and could still contribute to
the superpotential, provided it satisfy other necessary
constraints. In fact, an instanton wrapping this divisor does satisfy
the necessary constraints, and is responsible for generating the $10\,
10\, 5_H$ Yukawa coupling in a GUT model discussed in
\cite{Blumenhagen:2008zz,Cvetic:2010rq}. 

We remind that reader that while
it is necessary for an instanton to not intersect the D7 branes or to do
so at a $\bP^1$ in order to have an uncharged superpotential contribution,
the latter case is not sufficient. Specifically, if it intersects a $D7$ brane
at a $\bP^1$, one must explicitly verify whether or not it has charged modes.
In the case of the GUT model studied on this manifold, 
D7 worldvolume fluxes were turned on for the sake of chirality,
which greatly affects the charged mode calculation.

\subsubsection*{Calabi-Yau hypersurface in $dP_1\times \bF_0$}

We now present a geometry which is qualitatively different from the
geometries that we have discussed so far. Up to this point, we have
discussed weighted projective spaces and elliptic fibrations, where
the latter can be seen as the toric data for some base $\cB$ augmented
by a copy of $\bP_{231}$. Though this choice was natural for an
elliptic fibration in Weierstrass form, we now present an example of
an elliptically fibered Calabi-Yau threefold with a factorizing
intersection form where the toric ambient space does not explicitly
contain a copy of $\bP_{231}$. Instead of $\bP_{231}$, one could
augment the base $\cB$ by any of the two-dimensional toric varieties
represented by the sixteen reflexive two-dimensional polyhedra, as
each admits a Calabi-Yau onefold hypersurface, i.e. an elliptic curve.

\TABLE{
\begin{tabular}{c|cccccccc}
	& $s$ &$t$ &$u$ &$v$ &$w$ &$x$ &$y$ &$z$ \\  \hline
    $M$ & $1$ & $1$ & $1$ & $0$ & $0$ & $0$ & $0$ & $0$ \\
    $N$ & $0$ & $1$ & $0$ & $1$ & $0$ & $0$ & $0$ & $0$ \\
	$O$ & $0$ & $0$ & $0$ & $0$ & $1$ & $1$ & $0$ & $0$ \\
	$P$ & $0$ & $0$ & $0$ & $0$ & $0$ & $0$ & $1$ & $1$
\end{tabular}
\label{table:dP1P1P1}
\caption{GLSM charges for the 4D toric ambient space $dP_1\times (\bP_1 \times\bP_1)$.}
} For the sake of concreteness, we consider the four-dimensional toric
variety $\cA=dP_1\times \bF_0$ with $SRI=\langle su,tv,wx,yz\rangle$
and whose GLSM charges are given in table \ref{table:dP1P1P1}. The
intersection three form on the Calabi-Yau hypersurface $X$ is given by
\begin{equation}
\label{eqn:dP1P1P1iForm}
I_X = 2MN(O+P)-2N^2(O+P) + 2MOP+NOP.
\end{equation}
The intersection form does not factorize because $h^{1,1}(X)=4\ne
h^{1,1}(dP_1) + 1 = 2+1$. The fact that conditions on linear
equivalence and the Stanley-Reisner ideal are satisfied guarantees
that there are no non-zero triple intersections of the pullbacks of
the $dP_1$ divisors to the ambient space, but here either $O$
\emph{or} $P$ must participate in every non-zero triple intersection.
Thus, we explicitly see the effect of violating the conditions on K\" ahler
moduli necessary for factorization.

We wish to check whether or not this Calabi-Yau threefold admits
an elliptic fibration. This occurs \cite{oguiso} when there is an
effective divisor $D$ such that:
\begin{enumerate}
\item $D\cdot \cC\geq 0$ for all curves $\cC\in X$
\item $D^3=0$
\item $D^2\cdot D_i\neq 0$ for some divisor $D_i$ of $X$.
\end{enumerate}
We take $D=\sum_{i\in\{s,t,u,v\}} \pi^{-1}(C_i) = 3M+2N$.  It is clear
from the intersection form that $D^3=0$, and moreover it is
straightforward to calculate $D^2P=16\ne 0$, so that second and third
conditions are satisfied. We must also check that $D\cdot C \ge0$ for
all curves $C$ in the Mori cone. Since in our case elements of the
Mori cone can be written as a linear combination of $D_i\cdot D_j$, it
is a sufficient (but not necessary) condition to check that $D\cdot
D_i \cdot D_j \ge 0 \,\,\,\,\forall i,j \in \{s,t,u,v,w,x,y,z\}$.
Direct calculation shows that this is satisfied, and thus this
Calabi-Yau threefold admits an elliptic fibration. This example shows
explicitly that not all elliptic Calabi-Yau threefolds have a
factorizing intersection form.

One might also think that we have presented a boring example, where
the elliptic fibration is a trivial since the ambient toric variety is
a product space. However, it is not necessarily true that the
triviality of an ambient space fibration descends to the fibration of
some hypersurface. As a counterexample, $dP_1$ can be embedded in the
product space $\bP_2\times \bP_1$, as discussed in appendix
\ref{sec:equivariant-cohomology}. Similarly, $X$ is a non-trivial
elliptically fibered threefold, despite the fact that the toric
ambient space is a product space. This can be seen from the fact that a
trivial elliptic fibration has $\chi(\cB\times T^2) =
\chi(\cB)\chi(T^2)=0$, but $\chi(X)=-128$.

Finally, it is worth noting one more way in which this manifold is qualitatively
different from the others we have considered. In the examples where the toric
fibration was non-trivial, as in the case of the ambient spaces for Weierstrass
fibrations with the fiber a hypersurface in $\bP_{231}$, one could verify that $D_z$
was the expected base ($\bP_2$, $dP_2$, et cetera) by computing $K_{D_z}^2=\int_{D_z} c_1^2(T_{D_z})$
and $\chi(D_z)=\int_{D_z} c_2(T_{D_z})$ by the usual techniques. Here, though,
the fact that toric fibration is trivial guarantees that every non-zero triple
intersection of divisors must involve at least one of $\{M,N\}$ and at least
one of $\{O,P\}$, as is seen explicitly in the intersection three-form. For $D=oO+pP$ we have
$K_D^2=0$, and therefore no divisor of this form is a $dP_1$, as one might hope. 
The divisor $D_z$ in the previous examples also would not have been as expected had the toric fibration been
trivial.

\subsection*{Factorization without admitting an elliptic fibration}
We present one last example for the purpose of making another
qualitative point. So far, we have presented many examples of elliptic
threefolds with factorizing intersection form, as well an example of
an elliptic threefold where the intersection form does not
factorize. In this section we present an example from
\cite{Kachru:1995wm,Candelas:1993dm,Morrison:1996na} which is a
Calabi-Yau threefold that does factorize, but does not admit an
elliptic fibration. Our point in presenting this example, as well as
the last one, is to show that though elliptic threefolds furnish many
examples of manifolds with factorizing intersection form, not all
elliptic threefold factorize, and not all manifolds which factorize
are elliptic threefolds.

Consider the particular degree twelve Calabi-Yau hypersurface $X$ in
$\bP_{11226}$ which was studied in \cite{Kachru:1995wm} in the context
of heterotic-type II duality.  Blowing up a singular curve, the
intersection numbers on the manifold were calculated in
\cite{Candelas:1993dm} to be
\begin{equation}
\cI_X = H(4H^2-2HL)
\end{equation}
There are only two possible classes of divisors which might satisfy
the $D^3=0$ condition for an elliptic fibration: those that are
multiples of $L$ and those that are multiples of $3H-2L$. A curve $l$
was shown to have negative intersection with $D=3H-2L$ in
$\cite{Morrison:1996na}$, where this argument was explicitly made, and
therefore this particular $D$ does not satisfy the conditions for the
threefold to admit an elliptic fibration.  One can see also that
$L^2F=0$ for all divisors $F$, and therefore $D=L$ does not satisfy
the conditions for the threefold to admit an elliptic
fibration. Instead, $D=L$ satisfies similar conditions for $X$ to
admit a $K3$ fibration. So, $X$ is an example of a Calabi-Yau
threefold with factorizing intersection form which does \emph{not}
admit an elliptic fibration but instead admits a $K3$ fibration.

\section{Conclusions}

\subsection*{General conclusions}

In this paper we have discussed some computability features of the
physical problem of computing effective potentials in the
landscape. As we saw in sections~\ref{sec:systematics} and
\ref{sec:hilbert}, the translation of the problem into basic number
theoretic terms immediately illuminates rather general features of any
potential solution, and in fact suggests that the problem may be
algorithmically unsolvable in complete generality in the landscape.

If this observation turns out to be true, then it has important
consequences for the study of string vacua. First, more pragmatically,
this forces upon us a ``patchy'' description of the landscape: at any
point in time, we will have at our disposal a finite set of tools for
studying dynamics in the landscape, which apply only to some sets of
string vacua. The non-computability results above would then imply
that these tools are necessarily limited: either they do not give
exact results in finite time for all well-defined questions, or they
do not apply to the whole landscape, only to some particular classes
of vacua inside it. An example is our discussion in
section~\ref{sec:elliptic-fibrations}. There our method of attack was
solving the index formulas directly. We saw that algorithms exist for
doing so in some generality only for particular classes of spaces, in
particular threefolds with factorizing intersection form, which
include many elliptically fibered threefolds. In any case, no matter
how sophisticated our computational toolkit becomes, our computability
conjecture implies that there will always be vacua that are
inaccessible to the tools developed so far.

This last observation highlights a more conceptual difficulty arising
from our results. Optimistically, it may well happen that our vacuum
turns out to be one for which computational techniques are known. Or
more physically, if our arsenal of computational tools is large
enough, it may be always possible to find computable vacua that match
the observed phenomenology to the available accuracy.\footnote{This is
  somewhat similar to the distinction between real numbers and
  \emph{computable} real numbers that arose in
  section~\ref{sec:systematics}.} Nevertheless, even if we can find
such phenomenologically acceptable computable vacua, one of the
prominent features of string theory is that it can in principle ask
--- and answer --- questions about \emph{why} we ended up in our
particular vacuum. In other words, we are in principle able to
formulate dynamical questions about the landscape. If non-computable
vacua abound, and are in a sense generic, this would be telling us
that the computational model behind the landscape dynamics is
necessarily non-classical. This reinforces the lesson in
\cite{Denef:2006ad}, where it was found that an algorithmic approach
to the landscape yields NP-complete problems, and it is hard to
imagine how to overcome these in a semi-classical
framework.\footnote{A recent paper exploring related ideas is
  \cite{Williamson:2010ch}. There it is argued that some simple
  statements in a toy version of the landscape are ZFC-undecidable.}
This is necessarily an important and deep lesson about the dynamics on
the landscape.

\medskip

It could be, of course, that the computability conjecture in
section~\ref{sec:hilbert} is not true, and the effective potential for
an arbitrary string compactification is always exactly computable
algorithmically. If such an algorithm was found, we could turn the
previous discussion on its head, and by formulating the right
questions $\cP$ we should be able to map out the landscape in great
detail. (This is another reason why we believe that such an
algorithmic procedure does not exist.) This would also have
interesting repercussions in number theory, providing us with a
semi-universal method for solving large classes of very complicated
diophantine equations.

\subsection*{Extension of the results to other corners of the
  landscape}

The discussion in this paper is just a first step, and much remains to
be done. We tried to be as inclusive as possible, but also precise in
our statements. This leaves room for both expanding the discussion of
computability to quantities other than non-perturbative F-terms in
large volume type IIB, and for figuring out new classes of backgrounds
where (subsets of) the F-terms are exactly computable. We briefly
commented on the natural extension to F-theory in
section~\ref{sec:F-theory}, but many classes of models other than
large volume IIB and F-theory are routinely used in string theory, and
it would be rather interesting to understand their computability
structure. Let us briefly discuss some classes of string vacua for
which we expect similar ideas to apply.

We managed to relate the problem of computing non-perturbative effects
to number theory because in IIB/F-theory all necessary computations
reduce to algebraic geometry, and in the particular case of complete
intersections in toric varieties we could show explicitly the
computability structure of the resulting problem in algebraic
geometry. The tool of choice in heterotic string model building is
generically also algebraic geometry, and thus the study of
non-perturbative effects on complete intersections in toric varieties
should also have a simple algorithmic characterization there.

Another class of problems that we expect admits a clear computational
description is branes located at toric singularities. In this case the
description is rather combinatorial, using dimer model techniques
\cite{Franco:2005rj} (see \cite{Kennaway:2007tq,Yamazaki:2008bt} for
reviews).\footnote{Abelian orbifolds are particular cases of toric
  singularities which also admit a CFT treatment. See for example
  \cite{Bianchi:2007fx,Argurio:2007vqa,Bianchi:2007wy,Camara:2007dy,Camara:2008zk,Ibanez:2008my,Angelantonj:2009yj,Bianchi:2009bg,Billo':2010bd,Camara:2010zm}
  and references therein for some recent papers discussing D-brane
  instantons from CFT.} The effects of D-brane instantons on toric
singularities can be described systematically in terms of
combinatorial quantities
\cite{Franco:2007ii,Ibanez:2007tu,Forcella:2008au}. Use of mirror
symmetry then allows us to extend the dimer model discussion to the
context of intersecting branes in IIA \cite{Feng:2005gw}. As a side
remark, since we are dealing with holomorphic quantities one can
sometimes connect the large volume approach in this paper directly to
the dimer model construction simply by moving in K\"ahler moduli space
\cite{Hanany:2006nm}, in these cases the discussion of this paper
applies unchanged (although it will require some translation between
the large and small volume descriptions).

\subsection*{Further developments from the number-theoretic
  perspective}

Our discussion of the number theoretical aspects of the problem was
somewhat superficial. There are a number of constructions in which
non-computability can be proven, and perhaps an embedding of some of
these in string theory can be found. This would definitively show that
the landscape is not classically computable. Also, one could devise or
adapt methods for solving the diophantine equations arising from index
formulas in backgrounds other than those Calabi-Yau threefolds with
factorizing intersection form. As an example of a class of diophantine
equations that have been well studied in the mathematics community we
would like to highlight elliptic curves. That is, imagine that we are
interested in studying instantons contributing to the superpotential,
and in computing the necessary condition $\chi(D,\cO_D)=1$ we obtain
an equation of the form:
\begin{align}
  \label{eq:integer-elliptic-curve}
  y^2+a_1xy + a_3 y - x^3 - a_2 x^2 - a_4 x - a_6 = 0 
\end{align}
here the coefficients $a_i$ are integers which depend on the geometry
of the Calabi-Yau, and $x,y$ are also integers parameterizing the
divisor wrapped by the instanton. We are taking $D=xD_x+yD_y$, with
$D_x,D_y$ two basis divisors of the
Calabi-Yau. Equation~\eqref{eq:integer-elliptic-curve} is in fact the
equation for an elliptic curve over the integers, and it is known that
there are generically just a finite number of integer points $(x,y)$
on~\eqref{eq:integer-elliptic-curve} \cite{smart1998}. From the point
of view of this paper, if we had a procedure for obtaining this finite
set of points we would have solved the problem of computing
superpotentials in this background, since checking the spectrum of
zero modes over the finite set of solutions
of~\eqref{eq:integer-elliptic-curve} will take a finite time. We are
not aware of a general procedure for solving elliptic equations over
the integers, but there are known algorithms that work for large
numbers of elliptic equations. We refer the reader to \cite{smart1998}
for a very readable review of the relevant techniques that also
discusses some other classes of potentially relevant diophantine
equations.

\acknowledgments

We would like to acknowledge interesting discussions with Lara
Anderson, Matthew Ballard, Andres Collinucci, and Denis Klevers. We
gratefully acknowledge the hospitality of the KITP during the Strings
at the LHC and in the Early Universe program for providing a
stimulating environment during the initial stages of this
work. I.G.E. thanks N. Hasegawa for kind support and constant
encouragement. This research was supported in part by the National
Science Foundation under Grant No. NSF PHY05-51164, DOE under grant
DE-FG05-95ER40893-A020, NSF RTG grant DMS-0636606, the Fay R. and
Eugene L. Langberg Chair, and the Slovenian Research Agency (ARRS).
\appendix

\section{$\bZ_2$ equivariant line bundle cohomology}
\label{sec:equivariant-cohomology}

\subsection{The equivariant holomorphic genus}

In addition to the arithmetic genus, there is another set of necessary
conditions on contributing instantons coming from Lefschetz's
equivariant genus. This is a version of the holomorphic genus that
takes into account a possible orientifold involution $\sigma$:
\begin{align}
  \chi^\sigma(\cM, E) = \sum_i (-1)^i\left(H^{(0,i)}_+(\cM, E) -
    H^{(0,i)}_-(\cM,E)\right)
\end{align}
where $E$ is the bundle (or sheaf) defined over the manifold $\cM$
whose genus we are interested in (in general $\cM$ is the instanton
worldvolume, and $E$ is the trivial sheaf $\cO_\cM$), and
$H^{(0,i)}_\pm(\cM,E)$ are the even and odd (under $\sigma$)
cohomologies of $E$. The Lefschetz equivariant genus formula then
states that:
\begin{align}
  \label{eq:Lefschetz-equivariant-index}
  \chi^\sigma(\cM, E) = \int_{M^\sigma} \ch_\sigma(E) \frac{\Td(TM^\sigma)}{\ch_\sigma(\wedge_{-1}\ov{NM^\sigma})} \,\,\,\,\, .
\end{align}
The various elements entering in this formula require some
explanation.\footnote{We would like to thank A.~Collinucci for
  illuminating discussions on
  eq.~\eqref{eq:Lefschetz-equivariant-index}.} We have denoted by
$M^\sigma$ the fixed locus in $M$ under the involution
$\sigma$. Generically, this fixed locus decomposes as a sum of
connected components $M^\sigma=\cup C^\sigma$. The different
components can have different dimensionality, as we will see
explicitly in examples below.

$TM^\sigma$ and $NM^\sigma$ denote the tangent and normal
bundles to the fixed locus, respectively, and $\Td(TM^\sigma)$ denotes the Todd
class of $M^\sigma$:
\begin{align}
  \Td(TM^\sigma) = 1 + \frac{1}{2}c_1 + \frac{1}{12}(c_1^2+c_2) +
  \frac{1}{24}c_1c_2+\ldots
\end{align}
where $c_1$ and $c_2$ are the first and second Chern classes of $TM^\sigma$.

For a vector bundle $F$, $\wedge_{-1}F$ denotes the (formal)
alternating sum of antisymmetric powers of $F$. As an example, taking
$F=\cL_1\oplus \cL_2 \oplus \cL_3$, with $\cL_i$ line bundles, we have that:
\begin{align}
  \wedge_{-1}F = \cO - \cL_1\oplus \cL_2\oplus \cL_3 + (\cL_1\otimes \cL_2)
  \oplus (\cL_1\otimes \cL_3) \oplus (\cL_2\otimes \cL_3) - \cL_1\otimes
  \cL_2\otimes \cL_3.
\end{align}

Finally, $\ch_\sigma$ denotes an equivariant Chern character, in the
following sense. Consider a line bundle $\cL$. Generically, different
sections of $\cL$ transform with different signs under $\bZ_2$, but,
as it turns out, for each component $C^\sigma$ of the fixed locus
$M^\sigma$ the non-vanishing sections transform with a definite sign
$s_\sigma(C^\sigma,\cL)=\pm 1$. For each fixed component $C^\sigma$ we
thus define $\ch_\sigma(\cL) = s_\sigma(C^\sigma,\cL)\cdot\ch(\cL)$. A
simple way to understand this for toric varieties is the
following. Generically, $C^\sigma$ is only fixed after applying an
appropriate $\bC^*$ (gauge) transformation. This means that, on the
fixed locus $C^\sigma$, the $\bZ_2$ equivariant action embeds in the
gauge group of the GLSM. By definition, all sections of $\cL$
transform with the same sign under gauge transformations.  We will use
this viewpoint below to our advantage, as it provides an efficient way
of computing $s_\sigma(C^\sigma,\cL)$.

\medskip

We will provide many examples of the use of this formula in the next
section, when we discuss equivariant line bundle cohomology. From the
point of view of this paper, the equivariant holomorphic genus gives
an additional necessary constraint which is easily checked with the
tools at hand. For example, $H_+^{(0,0)}(D,\cO_D)$ should be the only
non-vanishing cohomology in a IIB compactification without flux if an
instanton on $D$ is to contribute to the superpotential, and therefore
we need to have that
\begin{align}
  \chi^\sigma(D,\cO_D) = 1.
\end{align}
As an illustration of the techniques, we will check the index formulas
for the $O(1)$ instanton candidate found in the elliptic fibration over $dP_2$
studied in section~\ref{sec:further examples}, which wraps
$P=D_8$.

Below, when writing the expressions, we will assume that the instanton
cycle is not pointwise invariant under the orientifold action, but
only curves and (possibly) isolated points inside the instanton are
invariant. In the case where the whole instanton is invariant, we have
an ordinary gauge instanton for a $USp(2N_c)$ or $SO(N_c)$ gauge group,
which will only contribute to the superpotential if $N_c\geq 0$, and $N_f=N_c$
or $N_f=N_c-3$, respectively \cite{Intriligator:1995id,Akerblom:2006hx}.

We will also mostly focus on the case of an instanton wrapping a
divisor inside a Calabi-Yau threefold. In this case the general
formulas simplify, and the equivariant genus receives contributions
only from isolated fixed points (denoted as $O3$ below) and
fixed curves (denoted by $M^\sigma$ below). Let us analyze these in
turn, starting with isolated fixed points.

\subsubsection*{Isolated fixed points}

In this case, the general Lefschetz formula gives for each fixed point
$O3$ intersecting $M$:
\begin{align}
  \label{eq:equivariant-genus}
  \begin{split}
    \chi^\sigma(M,\cL) & = \int_{O3}\ch_\sigma(\cL)
    \frac{\Td(TO3)}{\ch_\sigma(\wedge_{-1}\ov N_{O3})}\\
    & = \int_{O3} \frac{s_\sigma(O3,\cL)}{\ch_\sigma(\cO - \ov
      N_{O3} +
      \wedge^2 \ov N_{O3})} \\
    & = \int_{O3} \frac{s_\sigma(O3,\cL)}{\ch(\cO +
      \ov N_{O3} + \wedge^2 \ov N_{O3})}\\
    & = \int_{O3} \frac{s_\sigma(O3,\cL)}{1 + \rk(N_{O3})
      + \rk(\wedge^2
      N_{O3})}\\
    & = \int_{O3} \frac{s_\sigma(O3,\cL)}{1+2+1} =
    \frac{1}{4} s_\sigma(O3,\cL)
  \end{split}
\end{align}
where we have used that the normal bundle is antisymmetric under the
involution, that the normal bundle to a point inside a surface has
rank two, and that its second antisymmetric power has rank
one.\footnote{Locally, write the normal bundle as $N_{O3}=L_1\oplus
  L_2$, and thus $\wedge^2 N = L_1\otimes L_2$.} The final formula is
thus remarkably simple: each point where the divisor under
consideration intersects an isolated O3 contributes $\pm 1/4$ to the
equivariant genus, depending on the $\bZ_2$ character of the fixed
point.

In the example which is an elliptic fibration over $dP_2$, we have O3 planes at $x_3 = x_4 = x_8 = 0$ and at $x_3
= x_4 = x_7 = 0$. The first point obviously intersects $D_8$, and
using the intersection form~\eqref{eq:dP2-fibration-IX} on the
Calabi-Yau we have $D_3\cdot D_4 \cdot D_8 = +1$. The second O3 can be
seen not to intersect $D_8$: although $D_3\cdot D_4\cdot D_7$ does
intersect $D_8$ in the ambient space, this intersection point is
outside the Calabi-Yau hypersurface. One easy way to see this is
noting that the Calabi-Yau hypersurface has class $3D_6$ and
$x_6x_7x_8$ is in the Stanley-Reisner ideal. So in our there is one $O3$ plane
which contributes $\frac{1}{4}$ to $\chi^\sigma$ (the
character is $+1$ since we are dealing with the trivial bundle
$\cO$).

\subsubsection*{Fixed curves}

Fixed curves $M^\sigma$ in $M$ also contribute in a simple way to the
equivariant genus (see the appendix of \cite{Blumenhagen:2010ja} for a
recent similar discussion in the context of F-theory). Specializing to
the case of $\cL$ being a line bundle and
expanding~\eqref{eq:Lefschetz-equivariant-index}, we get
\begin{align}
  \begin{split}
    \chi^\sigma(\cM,\cL) &=
    \frac{s_\sigma(M^\sigma,\cL)}{2}\int_{M^\sigma}(c_1(\cL) +
    \frac{1}{2}c_1(TM))\\
    &= \frac{s_\sigma(M^\sigma,\cL)}{2}\int_{M^\sigma} (c_1(\cL) -
    \frac{1}{2}[M]) \,\,\,\,\, .
  \end{split}
\end{align}
Here $[M]$ should be understood as the Poincare dual two-form to $M$ in
the ambient Calabi-Yau $X$, and $TM$ is the tangent bundle to
$M$. This expression admits a simple expression in terms of
intersection numbers:
\begin{align}
  \chi^\sigma(\cM,\cO) = \frac{s_\sigma(M^\sigma,\cL)}{2} \left(\cL -
  \frac{1}{2}M\right)\cdot M\cdot \Pi_{O7}
\end{align}
where $\Pi_{O7}$ denotes the cycle wrapped by the orientifold planes,
and we abuse notation by also writing $\cL$ for the divisor associated
to the line bundle $\cL$.

For our particular example, we have
\begin{align}
  \begin{split}
    \chi^\sigma(D_8,\cO_{D_8}) &= -\frac{1}{4} D_8^2\cdot (D_2+D_5)\\
    & = -\frac{1}{4}D_8^2\cdot (D_1+2D_5) = -\frac{1}{4}(P^2\cdot M +
    2 P^2\cdot N) = \frac{3}{4}
  \end{split}
\end{align}
using the intersection form~\eqref{eq:dP2-fibration-IX}. Hence, adding
all contributions we have that
\begin{align}
\chi^\sigma(D_8) = 1,
\end{align}
so that this instanton satisfies the necessary condition for
superpotential contribution coming from Lefschetz's equivariant genus.

\subsection{Line bundle cohomology}

In the context of Calabi-Yau spaces constructed as hypersurfaces in
toric varieties, one can in fact compute explicitly the equivariant
structure of sheaf cohomology, and thus, in particular, the exact
spectrum of neutral zero modes on any instanton. The basic tool we
will need in order to do this is to compute the action of a $\bZ_2$
involution on the cohomology of line bundles on the ambient toric
variety. One can then use a Koszul sequence to project this down to
equivariant instanton cohomology, as we discuss in
section~\ref{sec:Koszul}.

Consider a toric space $\cA$. One can compute line bundle cohomology
for a line bundle $\cL$ by computing the \v Cech cohomology of a
particular complex built from the local sections of $\cL$. We will not
need to review the details of the construction here, and we refer the
reader instead to the appendix of \cite{Cvetic:2010rq} for a concise
review, and \cite{CLS:ToricVarieties} for a more systematic
exposition. The only point we need from that discussion is that, given
a monomial $s=x_1^{m_1}\cdots x_n^{m_n}$ describing a local section of
$\cL$, its contribution to the cohomology can be computed from its
behavior on the intersections of open sets in $\cA$ (in the sense of
whether it is a well defined section in a particular patch or not).

Consider now an involution acting on the coordinates of the GLSM for
$\cA$ as $x_i\mapsto (-1)^{s_i} x_i$, with $s_i$ arbitrary integers. This
transformation acts on a well defined way on the local sections,
namely:
\begin{align}
  \label{eq:section-Z2}
  s = x_1^{m_1}\cdots x_n^{m_n} \mapsto (-1)^{s_1m_1 + \ldots +
    s_nm_n}\,x_1^{m_1}\cdots x_n^{m_n}.
\end{align}
Notice that this transformation (trivially) does not change the
behavior of the local sections on the intersections of opens, in the
sense that all local sections in non-equivariant cohomology remain
local sections in equivariant cohomology, and the differential maps
are unaffected. Thus, the \v Cech complex determining the contribution
of this local section to the sheaf cohomology remains unchanged. So we
can understand the map~\eqref{eq:section-Z2} as giving the behavior of
the $\bZ_2$ involution on the sheaf cohomology of $\cA$. It is clear
that this recipe gives the right results for global sections, and
since the \v Cech complex is invariant we believe that it also gives
the correct result for local sections corresponding to higher sheaf
cohomologies. We have checked that this is true in a large number of
examples, some of which we present below.

In order to perform the enumeration of contributing local sections
efficiently it is very convenient to use the algorithm described in
\cite{Blumenhagen:2010pv}, and proven in
\cite{Jow:2010,Rahn:2010fm}. In this language the basic idea is that
the $\bZ_2$ action on each element of the line bundle cohomology is
given by the action on the representative rationom. Let us illustrate
how this works in a number of simple but illuminating examples.

\subsubsection*{Projective line: $\bP^1$}

The simplest example that we can consider is $\bP^1$, with coordinates
$(x_0,x_1)$. There is a single $\bZ_2$ involution of this space to
consider, namely $(x_0,x_1)\to (-x_0,x_1)$. Consider a line bundle
$\cO(k)$, with $k\geq 0$ for simplicity. The cohomology of this line
bundle comes only from its global sections: $H^n(\bP^1,\cO(k)) = 0$
for $n>0$, while $H^0(\bP^1,\cO(k))=k+1$. The sections are given by
the monomials:
\begin{align}
  x_0^n x_1^{k-n}\qquad \text{with }\, 0 \leq n \leq k
\end{align}
The inherited $\bZ_2$ action on these sections is then simply:
\begin{align}
  \label{eq:P1-sections-transform}
  x_0^n x_1^{k-n} \to (-1)^n x_0^n x_1^{k-n}
\end{align}
which induces the corresponding $\bZ_2$ action on $H^0(\bP^1,\cO(k))$.

In this simple case the result that we obtain is rather trivially
true, since the bundle is generated by its sections, but let us
double-check it by computing Lefschetz's holomorphic genus for the
line bundle. It is given by:
\begin{align}
  \label{eq:lefschetz-formula}
  \chi^\sigma(M,\cL) = \int_{\Phi}
  \frac{\ch_\sigma(\cL) \Td(T\Phi)}{\ch_\sigma(\wedge_{-1}\ov N\Phi)}
\end{align}
where we have denoted by $\Phi$ the fixed locus of the $\bZ_2$ action
$\sigma$, and the irreducible components of $\Phi$ by $\Phi_i$. In the
case under study $\sigma$ has two fixed points, $\Phi_1=(0,1)$ and
$\Phi_2=(0,1)$, and the integral splits into two components:
\begin{align}
  \label{eq:P1-index}
  \chi^\sigma(M,\cL) = \int_{(0,1)}\frac{1}{2} +
  \int_{(1,0)}\frac{(-1)^k}{2} = \sum_{i=0}^k (-1)^i.
\end{align}
The relative factor of $(-1)^k$ appears since the character of the
line bundle depends on the fixed component under consideration, as
explained above. It can be efficiently computed as follows: $\Phi_i$
may not actually be fixed under the $\bZ_2$ action $\sigma$ as a
subspace of the ambient $\bC^2=\{x_0,x_1\}$ (i.e., before imposing the
$\bC^*$ gauge invariance), and will only become a fixed locus once we
quotient $\bC^2$ by $\bC^*$. In other words, one may need to combine
$\sigma$ with some $\bC^*$ action $g_{\Phi_i}=\pm 1$ in order to leave
$\Phi_i$ fixed in the ambient space. A bundle with charge $k$ under
$\bC^*$ will then have the character $g_{\Phi_i}^k$. In our case, we
have that $(0,1)$ is fixed by itself under $\sigma$, so its associated
character is 1, while $(1,0)$ requires a $-1$ action on the covering
space, so its character is $(-1)^k$. This discussion generalizes
straightforwardly to higher dimensions and multiple $\bC^*$
symmetries, and we just quote the relevant results below.

\subsubsection*{Projective plane: $\bP^2$}

For our next example we move one complex dimension higher, and
consider $\bP^2$, parameterized by $(x_0,x_1,x_2)$. The $\bZ_2$
orientifold action $\sigma$ to consider in this case is given by:
\begin{align}
  (x_0,x_1,x_2) \to (-x_0,x_1,x_2)
\end{align}

Let us again consider for simplicity a bundle of the form $\cO(k)$,
with $k>0$. This bundle is ample, i.e. generated by its sections, and
the sections can be described as monomials of the form:
\begin{align}
  x_0^ax_1^bx_2^{k-a-b}\qquad\text{with }\, a\geq 0, b\geq 0, k\geq a+b
\end{align}
The action of $\sigma$ is thus given by:
\begin{align}
  \label{eq:P^2-equivariant}
  x_0^ax_1^bx_2^{k-a-b} \to (-1)^a x_0^ax_1^bx_2^{k-a-b}
\end{align}
inducing the corresponding action on the cohomology.

Let us check this against the result from Lefschetz's theorem. In this
case the fixed point set consists of the curve $x_0=0$, and the point
$(1,0,0)$. By a simple application of~\eqref{eq:lefschetz-formula} we
obtain:
\begin{align}
  \chi^\sigma(\bP^2, \cO(k)) = \left(\frac{1}{2}k + \frac{3}{4}
  \right) + \frac{(-1)^k}{4}
\end{align}
which can be easily seen to agree with the result obtained
from~\eqref{eq:P^2-equivariant}.

\subsubsection*{Lifting the geometric action to an action on the
  bundle}

Before proceeding to more involved examples, let us describe an
important subtlety that we have ignored in the previous examples,
namely, the fact that a single geometric action can have multiple
lifts to the complete line bundle. In our context, which lift we
consider can be encoded in the way we describe the $\bZ_2$
action. Consider for example $\bP^2$, described above. We took the
involution to be given by:
\begin{align}
  \sigma: (x_0,x_1,x_2) \to (-x_0,x_1,x_2)
\end{align}
but due to the $\bC^*$ symmetry of $\bP^2$, an equivalent description
of this geometric action is
\begin{align}
  \tau: (x_0,x_1,x_2) \to (x_0,-x_1,-x_2).
\end{align}

This ambiguity in fact encodes the two possible lifts of the geometric
action on $\bP^2$ to the bundle. Applying the same recipe as before,
but now with $\tau$ instead of $\sigma$, we obtain that the sections
of the bundle transform as:
\begin{align}
  \tau^*: x_0^a x_1^b x_2^{k-a-b} \to (-1)^{k-a} x_0^a x_1^b x_2^{k-a-b} \,\,\,.
\end{align}
That is, the action on each section gets multiplied by $(-1)^k$ with
respect to~\eqref{eq:P^2-equivariant}. It is easy to check that with
the prescription for the characters given above, the Lefschetz index
reproduces this result: the fixed components $x_0=0$, $x_1=x_2=0$ now
have characters $(-1)^k$ and $1$ respectively, and thus:
\begin{align}
  \chi^\tau(\bP^2, \cO(k)) = (-1)^k \chi^\sigma(\bP^2, \cO(k)).
\end{align}
From these arguments we see the need to define an action on the bundle
in addition to the action on the geometry. In the mathematical
literature this is commonly known as introducing an equivariant
structure on the bundle. Let us briefly review this standard
discussion.

Consider a vector bundle $V \stackrel{\pi}{\rightarrow} X$ and an
action of a discrete group $G$ on $X$. If for each $g\in G$ there
exists a bundle morphism $\phi_g:V\rightarrow V$ such that the diagram
\begin{equation}
  \begin{array}{lllll}
    &V&\stackrel{\phi_g}{\longrightarrow}&V&\\
    \pi &\downarrow&&\downarrow&\pi\\
    &X&\stackrel{g}{\longrightarrow}&X&
  \end{array}
  \label{invalt}
\end{equation}
commutes for all $g\in G$, then the bundle morphisms define an
\emph{invariant structure} on $V$. If the morphisms also satisfy the
relation
\begin{equation}
  \phi_g \circ \phi_h = \phi_{gh}
\end{equation}
for all $g,h\in G$, then the morphisms define an $\emph{equivariant
  structure}$ on $V$.  An equivariant structure induces a map on
sections. Requiring that the diagram
\begin{equation}
  \begin{array}{lllll}
    &V&\stackrel{\phi_g}{\longrightarrow}&V&\\
    s &\uparrow&&\uparrow&s'\\
    &X&\stackrel{g}{\longrightarrow}&X&
  \end{array} .
  \label{invalts}
\end{equation}
commutes for all $g\in G$ gives a map $\Phi_g:\Gamma(X,V) \rightarrow
\Gamma(X,V)$ on sections defined by
\begin{equation}
  s'=\Phi_g(s)= \phi_g \circ s \circ g^{-1}.
\end{equation}
Using the fact that $\phi_g$ give an equivariant structure, it can be
shown that these maps obey the property $\Phi_g \circ \Phi_h =
\Phi_{gh}$.

Having discussed equivariant structures and their induced sections, we
would like to try to understand the above discussions in this
formalism. For the purposes of this paper, the group $G$ is that of
the holomorphic involution corresponding to the orientifold, so that
$G=\bZ_2$. In studying equivariant indices and cohomology, it is
natural to study the induced (possibly local) section $\Phi_g(s)$
corresponding to a $g\in \bZ_2$.  Representing the $\bZ_2$ action on
$\bP^2$ as $\bZ_2=\{1,\sigma\}$, the morphisms
\begin{align}
  \phi_1: p\times n \mapsto 1\cdot p\times n \qquad
  \phi_\sigma:p\times n\mapsto \sigma \cdot p\times n
\end{align}
act on $p\times n \in \bP^2 \times \bC$ and
give an equivariant structure on $\cO_{\bP^2}(k)$.  A generic section
$s$ is of the form $s:(x_0,x_1,x_2) \mapsto (x_0,x_1,x_2) \times x_0^a
x_1^b x_2^{k-a-b}$.  A corresponding induced section is
$s'=\Phi_{\sigma}(s) = \phi_{\sigma} \circ s \circ \sigma^{-1}$ so
that
\begin{align}
s':(x_0,x_1,x_2) \mapsto (x_0,x_1,x_2) \times (-1)^a x_0^a x_1^b x_2^{k-a-b} \equiv (-1)^a s,
\end{align}
and we see that the equivariant structure we have defined gives the
expected induced section.  Furthermore, it is trivial that $\Phi_1(s)
= \Phi_\sigma \circ \Phi_\sigma (s) = s$, and thus we have determined
all induced sections under the $\bZ_2$ equivariant
structure. Similarly, if we represent the $\bZ_2$ action on $\bP^2$ by
$\bZ_2=\{1,\tau\}$, and define $\phi_1$ and $\phi_\tau$ similarly, one
obtains $\Phi_1(s)=s$ and $\Phi_\tau(s) = (-1)^{k-a}s$, as obtained above.

The point of this discussion is that to study equivariant indices and
equivariant line bundle cohomology, one must specify an equivariant
structure in addition to the group action on the manifold. In the
language of string theory, this means that an equivariant structure is
part of the input data for a large volume type IIB orientifold
compactification, in addition to the Calabi-Yau manifold and the
holomorphic involution.

\subsubsection*{Blow-up of $\bP^2$ at one point}

\TABLE{
  \label{table:dP1-GLSM}
  \centering
    \begin{tabular}{c|cccc}
      & $x_1$ & $x_2$ & $x_3$ & $x_4$\\
      \hline
      $\bC^*_1$ & 1 & 1 & 1 & 0\\
      $\bC^*_2$ & 0 & 0 & 1 & 1
    \end{tabular}
  \caption{GLSM for $dP_1$.}
}
Let us analyze a further example that tests the previous results in a
slightly more non-trivial way: the blow-up of $\bP^2$ at one point,
also known as the first del Pezzo surface $dP_1$. We can represent this space
torically by the gauged linear sigma model (GLSM) in
table~\ref{table:dP1-GLSM}.

There are two possible $\bZ_2$ involutions that we can consider in
this space \cite{Blumenhagen:2008zz}. Let us focus on the one given
by
\begin{align}
  (x_1,x_2,x_3,x_4) \to (-x_1,x_2,x_3,x_4).
\end{align}
We consider a line bundle $\cO(m,n)$ of charge $(m,n)$ under
$\bC^*_{1,2}$, with $m\geq 0,n\geq 0$ for simplicity. The nonvanishing
cohomology of this bundle comes only from $H^i(dP_1,\cO(m,n))$, with
$i=0,1$. In particular, contributions to $H^0$ come from monomials
$\prod x_i^{a_i}$ with $a_i\geq 0$ and total charge $(m,n)$ under
$\bC^*_i$, and contributions to $H^1$ come from rationoms
\cite{Blumenhagen:2010pv} of the form:
\begin{align}
  \mathfrak{h} = \frac{T(x_3,x_4)}{x_1x_2 W(x_1,x_2)}
\end{align}
such that $T$ and $W$ are monomials of positive degree of the
respective variables, and the total charge of $\mathfrak{h}$ is
$(m,n)$. As before, the action of $\sigma$ on the cohomology can be
read simply from its action on the representative rationoms:
\begin{align}
  \prod x_i^{a_i} \to (-1)^{a_1}\prod x_i^{a_i},\qquad \mathfrak{h}\to (-1)^{1+\mathrm{ord}_1(W)}\mathfrak{h},
\end{align}
where $\mathrm{ord}_1(W)$ means the order $a$ of $x_1$ in
$W(x_1,x_2)=x_1^ax_2^b$.

The fixed point locus of $\sigma$ in this case is given by the curve
$x_1=0$, and the two points $x_2 = x_3 = 0$, $x_2 = x_4 = 0$. The
Lefschetz formula gives:
\begin{align}
  \chi^\sigma(dP_1,\cO(m,n)) = \frac{1}{2}(n+1) + \frac{(-1)^m}{4}
  + \frac{(-1)^{m+n}}{4},
\end{align}
which can be seen to agree with the action on the rationoms.

\subsubsection*{Another $\bZ_2$ quotient of $dP_1$}

Let us study the other possible $\bZ_2$ quotient of $dP_1$:
\begin{align}
  \sigma_2:(x_1,x_2,x_3,x_4) \to (x_1,x_2,x_3,-x_4)  
\end{align}
The discussion for the action on the rationoms proceeds analogously to
the previous case, with the resulting action being
\begin{align}
  \label{eq:dP_1-2 second action}
  \prod x_i^{a_i} \to (-1)^{a_4}\prod x_i^{a_i},\qquad \mathfrak{h}\to (-1)^{1+\mathrm{ord}_4(T)}\mathfrak{h}.
\end{align}

The fixed point locus consists of two disjoint curves: $x_3=0$ and
$x_4=0$. The Lefschetz formula is in this case
\begin{align}
  \chi^{\sigma_2}(dP_1,\cO(m,n)) =
  \frac{1}{2}\left(m-n+\frac{1}{2}\right) + \frac{(-1)^n}{2}\left(m +
    \frac{3}{2}\right),
\end{align}
which can easily be seen to give results consistent with the action on
the cohomology induced from~\eqref{eq:dP_1-2 second action}.

\subsubsection*{Blow-up of $\bP^2$ at three points}

Let us now consider the blow-up of $\bP^2$ at three points, also known
as the third del Pezzo surface $dP_3$. The GLSM model data for $dP_3$
is shown in table~\ref{table:dP3}.  \TABLE{
  \label{table:dP3}

  \centering
  \begin{tabular}{c|cccccc}
    & $x_1$ & $x_2$ & $x_3$ & $x_4$ & $x_5$ & $x_6$\\
    \hline
    $\bC^*_1$ & 1 & 1 & 1 & 0 & 0 & 0\\
    $\bC^*_2$ & 0 & 0 & 1 & 1 & 0 & 0\\
    $\bC^*_3$ & 0 & 1 & 0 & 0 & 1 & 0\\
    $\bC^*_4$ & 1 & 0 & 0 & 0 & 0 & 1
  \end{tabular}
  \caption{GLSM charges for $dP_3$.}
}

Consider a bundle $\cL$ with charges $q_i$ under the $\bC^*_i$, and a
$\bZ_2$ action $\sigma$ given by:
\begin{align}
  \label{eq:dP3-involution}
  \sigma: (x_1,x_2,x_3,x_4,x_5,x_6) \to (x_1,x_2,x_3,x_4,x_5,-x_6)
\end{align}
This action has a couple of fixed curves at $x_1=0$ and $x_6=0$, with
characteristic $(-1)^{q_4}$ and 1 respectively, and two fixed points
at $x_3=x_5 = 0$ and $x_2=x_4=0$, with characteristics
$(-1)^{q_1+q_3+q_4}$ and $(-1)^{q_1+q_2+q_4}$
respectively.\footnote{Looking at the GLSM only, one also obtains
  $x_2=x_3=0$ as a fixed locus, but it is in the Stanley-Reisner ideal
  of $dP_3$.} From here, the equivariant Lefschetz index is easily
calculated to be:
\begin{align}
  \begin{split}
    \chi^\sigma(dP_3,\cL) &= \frac{1}{2}\left(q_1 - q_4 +
      \frac{1}{2}\right) +
    \frac{(-1)^{q_4}}{2}\left(-q_1 + q_2 + q_3 + \frac{1}{2}\right)\\
    &+ \frac{(-1)^{q_1+q_3+q_4}}{4} + \frac{(-1)^{q_1+q_2+q_4}}{4}.
  \end{split}
\end{align}

In order to compute the action of the involution $\sigma$ on the
cohomology we proceed as before, and assign to each element of the
cohomology a sign given by the straight lift of the
action~\eqref{eq:dP3-involution} to the representative rationoms. In
the case of $dP_3$ we encounter for the first time the issue of a
given rationom contributing more than one element to the cohomology,
encoded in the appearance of non-trivial remnant cohomology in the
algorithm of \cite{Blumenhagen:2010pv}. By studying examples, we see
that the whole secondary (remnant) cohomology transforms in the same
way as the representative rationom, and thus one just needs to
multiply by the appropriate prefactor when calculating equivariant
indices. This agrees with the observation made at the beginning of
this section that orientifolds acting as a sign change do not act on
the \v Cech complex for a given local section.

\subsubsection*{A realistic example}

The previous examples are illuminating but not particularly
realistic. For our purposes we are interested in computing equivariant
line bundle cohomology on four complex dimensional toric ambient
spaces, rather than complex surfaces as we discussed above. Let us
briefly discuss the issues that arise in applying the formalism above
to a realistic example.\footnote{In dealing with realistic toric
  spaces we found the computer program SAGE \cite{sage} extremely
  useful, in particular in conjunction with its package for dealing
  with toric varieties \cite{toricsage}.}

\TABLE{
  \label{table:X1-GLSM}

  \centering
  \begin{tabular}{c|ccccccc}
    & $x_1$ & $x_2$ & $x_3$ & $x_4$ & $x_5$ & $x_6$ & $x_7$\\
    \hline
    $\bC^*_1$ & 1 & 1 & 1 & 0 & 1 & 0 & 0\\
    $\bC^*_2$ & 0 & 0 & 1 & 1 & 0 & 1 & 0\\
    $\bC^*_3$ & 0 & 1 & 0 & 0 & 1 & 0 & 1
  \end{tabular}

  \caption{GLSM data for $\cA$.}
} Consider the 4d toric variety $\cA$ described by
the GLSM data in table~\ref{table:X1-GLSM}. This space admits a smooth
triangulation described by the Stanley-Reisner ideal:
\begin{align}
  \cI(\cA) = \langle x_1x_3, x_1x_2x_5, x_3x_4x_6, x_4x_6x_7, x_2x_5x_7 \rangle
\end{align}
We take the $\bZ_2$ involution to be $\sigma:x_4\mapsto -x_4$. This obviously
leaves the $x_4=0$ locus invariant, and by using $\bC^*$ gauge
transformations it is easy to see that the following loci are also
fixed under the involution:
\begin{align}
  \begin{split}
    x_3 = x_6 = 0 & \qquad (-1)^n\\
    x_1 = x_6 = x_7 = 0 & \qquad (-1)^{n+m+p}
  \end{split}
\end{align}
We have listed next to each component of the fixed locus the character
of a line bundle $\cL = \cO(m,n,p)$ on it, with $(m,n,p)$ denoting as
usual the charges of the bundle under the $\bC^*$ symmetries of the
GLSM. There is formally another component of the fixed locus given by
$x_1=x_2=x_5=x_6=0$, but it belongs to the Stanley-Reisner ideal and
therefore we do not consider it further.

We see that in this case we have fixed loci of complex codimension 1,
2 and 3. Expanding~\eqref{eq:Lefschetz-equivariant-index} to the
relevant orders, we get the following expressions. For the codimension
1 surface $D_4:\{x_4=0\}$, there is contribution to the index given
by:
\begin{align}
  \label{eq:X1-I1}
  \begin{split}
    \delta\chi^\sigma(\cA,\cO(m,n,p)) &= \frac{1}{2}\int_{D_4} \ch(\cO(m,n,p))
    \frac{\Td(TD_4)}{1-\frac{1}{2}D_4 + {1}{4}D_4^2 -
      \frac{1}{12}D_4^3+\ldots} \\&= \frac{1}{48}(-4m^3 + 12m^2n - 12mn^2
    + 4n^3 + 12mp^2 - 8p^3\\&\phantom{=\frac{1}{48}(} - 6m^2 + 12mn - 6n^2 + 36mp - 12p^2 + 28m -
    4n + 20p + 27)
  \end{split}
\end{align}
The contribution from the codimension 2 surface $\cS:\{x_3=x_6=0\}$
is:
\begin{align}
  \label{eq:X1-I2}
  \begin{split}
    \delta\chi^\sigma(\cA,\cO(m,n,p)) & = \frac{(-1)^n}{4}\int_\cS
    \ch(\cO(m,n,p))\frac{\Td(T\cS)}{1 - \ch(\ov{D_3}\oplus\ov{D_6}) +
      \ch(\ov{D_3}\otimes\ov{D6})}\\
    & = \frac{(-1)^n}{8}(p^2 + 3p + 2)
  \end{split}
\end{align}
where we have denoted by $D_3$ and $D_6$ the line bundles associated
to the corresponding divisors. We also have a fixed curve
$\cC:\{x_1=x_6=x_7=0\}$, which gives a contribution:
\begin{align}
  \label{eq:X1-I3}
  \begin{split}
    \delta\chi^\sigma(\cA,\cO(m,n,p)) & = \frac{(-1)^{n+m+p}}{8}\int_\cC
    \ch(\cO(m,n,p)) \frac{\Td(T\cC)}{1-\frac{1}{8}([D_1] + [D_6] +
      [D_7])}\\
    & = \frac{(-1)^{n+m+p}}{48}(6p+9)
  \end{split}
\end{align}
Finally, and although we have no such case in our example, let us
mention for completeness that fixed points would give a contribution
of $\pm 1/16$ to the Lefschetz index. The Lefschetz equivariant index
in our example is thus the sum of~\eqref{eq:X1-I1}, \eqref{eq:X1-I2}
and \eqref{eq:X1-I3}.

Let us check in a simple example that the result of our prescription
on the rationoms agrees with the Lefschetz index. Consider the line
bundle $\cO(1,1,1)$. For this bundle only $H^0$ is non-vanishing. The
contributing sections with their sign under $\sigma:x_4\mapsto -x_4$
are:
\begin{align}
  \begin{matrix}
    x_3x_7 & \quad(+) &\qquad \qquad& x_1x_6x_7& \quad(+)\\
    x_1x_4x_7 & \quad(-) && x_5x_6& \quad(+)\\
    x_2x_6 & \quad(+)&&x_4x_5 & \quad(-)\\
    x_2x_4&\quad(-)&&&
  \end{matrix}
\end{align}
From here, it is clear that $h^0_+ - h^0_-=1$. Substituting $m=n=p=1$
in the expression for the Lefschetz equivariant index found above
agrees with this result.

\subsection{Koszul resolution}
\label{sec:Koszul}

In the previous section we have discussed how to compute equivariant
line bundle cohomology on toric varieties. Nevertheless, in physical
applications one is actually interested in equivariant line bundle
cohomology on varieties which are not toric, compact Calabi-Yau spaces
being a notable example. Luckily, most spaces of interest (denoted $X$
in what follows) can be embedded as complete intersections of
hypersurfaces in toric ambient spaces $\cA$, and one can carry over
the information obtained in the previous section to the subspace.

Our basic tool will be the Koszul complex:
\begin{align}
  0 \to N^* \xrightarrow{f} \cO_\cA \xrightarrow{r} \cO_X \to 0
\end{align}
where $N^*=\cO_\cA(-X)$ is the dual to the normal bundle of $X$ in
$\cA$, and we are assuming here for ease of exposition that $X$ is a
divisor in $\cA$ (the general expression is given for example in the
appendix of~\cite{Cvetic:2010rq}). The first map is multiplication by
the section $f=0$ defining $X$, and the second map is restriction to
$X$. For our applications, it will be convenient to tensor this exact
sequence with appropriate line bundles $\cO(D)$, giving:
\begin{align}
  \label{eq:Koszul-sheaves}
  0 \to \cO_\cA(D-X) \xrightarrow{f} \cO_\cA(D) \xrightarrow{r} \cO_X(D) \to 0
\end{align}
By the snake lemma, this exact sequence gives a long exact sequence in
cohomology:
\begin{align}
  \label{eq:Koszul-cohomologies}
  \begin{split}
    0 & \to H^0(\cA,\cO_\cA(D-X)) \to H^0(\cA,\cO_\cA(D)) \to
    H^0(X,\cO_X(D)) \to\\
    & \to H^1(\cA,\cO_\cA(D-X)) \to H^1(\cA,\cO_\cA(D)) \to
    H^1(X,\cO_X(D)) \to \ldots
  \end{split}
\end{align}
where the maps are the ones induced from \eqref{eq:Koszul-sheaves}. As with
any long exact sequence of abelian groups, this exact sequence can be split
into short exact sequences. In general, given a long exact sequences
of abelian groups:
\begin{align}
  \ldots \to A \to B \to C \to D \to E \to 0
\end{align}
there exists a $X$ such that both
\begin{align}
  \ldots \to A \to B \to C \to X \to 0
\end{align}
and
\begin{align}
  \label{eq:Koszul-short-split}
  0 \to X \xrightarrow{x} D \to E \to 0
\end{align}
are exact. In this way we can split any long exact sequence such
as~\eqref{eq:Koszul-cohomologies} into short exact sequences. Consider
for example the short exact sequence~\eqref{eq:Koszul-short-split}. We
have that
\begin{align}
  E \simeq \frac{D}{Im(X)},
\end{align}
where $Im(X)$ denotes the image of $X$ under $x$. Since $x$
is injective, $Im(X)$ can be thought of as an embedding of $X$ in
$D$.

Consider now the case of the Koszul
sequence~\eqref{eq:Koszul-cohomologies}, and for simplicity let us
just assume that $H^1(\cA,\cO_\cA(X-D))=0$. Therefore we do not have to split
the long exact sequence, since we already have the short exact
sequence
\begin{align}
  0 \to H^0(\cA,\cO_\cA(D-X)) \xrightarrow{f} H^0(\cA,\cO_\cA(D)) \to
  H^0(X,\cO_X(D)) \to 0.
\end{align}
and thus we have that
\begin{align}
  \label{eq:Koszul-isomorphism}
  H^0(X,\cO_X(D)) \simeq \frac{H^0(\cA,\cO_\cA(D-X))}{f^*(H^0(\cA,\cO_\cA(D)))} \,\,\, .
\end{align}
Recall that the map $f$ denotes multiplication by the equation
defining the hypersurface $X$, and here $f^*$ is its pullback to the
space of sections (we can also think of this as multiplying by an
specific section of $\cO_\cA(D)$). Notice that we have obtained that
the unknown cohomology group $H^0(X,\cO_X(D))$ can be expressed in
terms of known cohomologies in the ambient space. In order to study
systematically the representation of $H^0(X,\cO_X(D))$ under the
$\bZ_2$ action, it will be convenient to introduce some elementary
group theory.

Elements of line bundle cohomology groups transform in specific
representations of the $\bZ_2$ action, and we would now want to split
into irreps of $\bZ_2$, namely terms that transform with a plus sign
and terms that transform with a minus sign. Since $\bZ_2$ is a finite
group, this information is completely encoded in the group character:
\begin{align}
  \chi_g(R) = \Tr_R(g)
\end{align}
where $\Tr_R$ simply denotes the trace over the representation $R$. By
Schur orthogonality, the characters of the irreps of a finite group
are orthogonal under the inner product
\begin{align}
  \chi(R)\cdot \chi(S) = \frac{1}{|G|}\sum_g \chi_g^*(R) \chi_g(S)
\end{align}
where $\chi(R)=(\chi_1(R),\dots,\chi_{g_{|G|}}(R))$
and $|G|$ is the order of the group (2 for $\bZ_2$). This is evident in
our case, since
\begin{align}
  \chi_g(+) = (1,1),\qquad \chi_g(-) = (1,-1).
\end{align}
where we have denoted by $\pm$ the trivial and fundamental
representations of $\bZ_2$. From these expressions, if we have a
cohomology group $H$ decomposing as $H_{+}\oplus H_{-}$, we have that
\begin{align}
  \dim H_{\pm} = \frac{1}{2}\chi(H)\cdot (1,\pm 1).
\end{align}
So in order to obtain the relevant dimensions, we need to know the
character of $H$.

Let us come back to~\eqref{eq:Koszul-isomorphism}. Consider first the
case in which the map $f$ is invariant under the $\bZ_2$
involution. In this case, by the
isomorphism~\eqref{eq:Koszul-isomorphism} and some basic facts about
representation theory \cite{fulton-harris} (see also
\cite{Anderson:2009mh} for a similar recent discussion in the context
of the heterotic string) we have that:
\begin{align}
  \label{eq:character-rule}
  \chi_g(H^0(X,\cO(D))) = \chi_g(H^0(\cA,\cO_\cA(D))) -
  \chi_g(H^0(\cA,\cO_\cA(X-D))).
\end{align}
The other possibility is that $f$ takes a minus sign when we act with
$\bZ_2$ ($f=0$ is still invariant, of course). In this case the
inclusion map $f^*$ introduces an additional minus sign into
\eqref{eq:character-rule}, which is now given by:
\begin{align}
  \label{eq:signed-character-rule}
  \chi_g(H^0(X,\cO(D))) = \chi_g(H^0(\cA,\cO_\cA(D))) -
  g\cdot\chi_g((H^0(\cA,\cO_\cA(X-D)))).
\end{align}
where $g=\pm 1$. A similar discussion applies for more complicated
situations, with the net effect that one has to multiply
$H^i(\cA,\cO_\cA(X-D))$ by $g$ when using character addition
formulas such as~\eqref{eq:signed-character-rule}.

\subsubsection*{Example: $dP_1\subset \bP^2\times\bP^1$}

As an illustration of the method, let us discuss line bundle
cohomology for $dP_1$, now understood as a hypersurface in
$\bP^2\times\bP^1$. In particular, denoting as $(z_0,z_1,z_2|y_0,y_1)$
the coordinates of $\bP^2\times\bP^1$, $dP_1$ can be understood as any
hypersurface of the form $\sum c_{ik} z_iy_k=0$. For definiteness, let
us take $f=z_1y_1 + z_2y_0=0$ to be our chosen representative. Since
we will want to compare results with the results derived above for the
equivariant cohomology of $dP_1$, let us parameterize the $dP_1$ as
above in table~\ref{table:dP1-GLSM}, in terms of $x_1\ldots x_4$. In
these variables, it is easy to see that one explicit embedding is
given by:
\begin{align}
  \label{eq:dP1-embedding}
  \begin{split}
    z_0 = x_3&\qquad y_0 = x_2\\
    z_1 = x_2x_4 &\qquad y_1 = -x_1\\
    z_2 = x_1x_4
  \end{split}
\end{align}
Consider a $\bZ_2$ action on $\bP^2\times \bP^1$ of the
form:\footnote{Recall that in our conventions~\eqref{eq:P^2xP^1-Z_2}
  actually specifies an action on the bundle, not just on the
  geometry.}
\begin{align}
  \label{eq:P^2xP^1-Z_2}
  (z_0,z_1,z_2|y_0,y_1) \to (z_0,z_1,-z_2|-y_0,y_1)
\end{align}
This action leaves fixed a curve and two points inside the $dP_1$
hypersurface, and we can thus identify it with the first involution of
$dP_1$ studied above. Indeed, from the
embedding~\eqref{eq:dP1-embedding} we obtain an induced action on the
$dP_1$ sections:
\begin{align}
  \label{eq:dP1-induced-action}
  (x_1,x_2,x_3,x_4) \to (x_1,-x_2,x_3,-x_4).
\end{align}

As a further piece of information before going into the Koszul
resolution, we need to identify the divisors in the ambient space with
the divisors in the $dP_1$ hypersurface. This is easily done by
imposing that the induced intersection forms and the tangent bundle on
$dP_1$ agree in both bases. Denoting $D_1,D_2$ the hyperplanes of
$\bP^1$ and $\bP^2$ respectively, and $H,X$ the $(1,0)$ and $(0,1)$
divisors of $dP_1$ in the conventions of the previous section, we have
that $X=D_2-D_1$, $H=D_1$.

From here on it is just a matter of checking the formulas. Take for
example the divisor $\cO_{dP1}(3H+X)$. The only non-vanishing elements
of the induced cohomology are (we abbreviate $\bP^2\times \bP^1$ to
$\cA$):
\begin{align}
  0 \to H^0(\cA, \cO_{\cA}(D_1)) \to H^0(\cA, \cO_{\cA}(D_2+2D_1)) \to
  H^0(dP_1, \cO_{dP1}(3H+X))\to 0.
\end{align}
An easy computation using the techniques described above, or
alternatively a combination of more classical methods such as the
K\"unneth formula and Lefschetz's equivariant index theorem, gives:
\begin{align}
  \begin{split}
    \chi_g(H^0(\cA, \cO_{\cA}(D_1))) &= (2, 0)\\
    \chi_g(H^0(\cA, \cO_{\cA}(2D_1+D_2))) &= (9, 1)
  \end{split}
\end{align}
We have that $f=z_1y_1 + z_2y_0$ is invariant
under~\eqref{eq:P^2xP^1-Z_2}, so using~\eqref{eq:character-rule} we
deduce that:
\begin{align}
  \chi_g(H^0(dP_1, \cO_{dP1}(3H+X)) = (7,1)
\end{align}
which can be easily verified independently using the formulas in the
previous section. Notice that it is important when checking these
formulas to take the proper restriction of the bundle $\bZ_2$ action,
this is the one given in~\eqref{eq:dP1-induced-action}.

As another quick example, consider the $\bZ_2$ action leaving two
curves on $dP_1$ fixed:
\begin{align}
  (x_1,x_2,x_3,x_4) \to (x_1,x_2,x_3,-x_4).  
\end{align}
It is easy to see that this lifts to:
\begin{align}
  \label{eq:P^2xP^1-Z_2b}
  (z_0,z_1,z_2|y_0,y_1) \to (z_0,-z_1,-z_2|y_0,y_1)
\end{align}
Taking the same bundle as before, the characters are now:
\begin{align}
  \begin{split}
    \chi_g(H^0(\cA, \cO_{\cA}(D_1))) &= (2, 2)\\
    \chi_g(H^0(\cA, \cO_{\cA}(2D_1+D_2))) &= (9, -3)
  \end{split}
\end{align}
We now have that $f=z_1y_1 + z_2y_0$ changes sign under the $\bZ_2$
action~\eqref{eq:P^2xP^1-Z_2b}, and thus
using~\eqref{eq:signed-character-rule} we have that:
\begin{align}
  \chi_g(H^0(dP_1, \cO_{dP1}(3H+X)) = (7,-1).
\end{align}
which can also be checked independently using the techniques of the
previous section.

\subsection{Permutation orientifolds}

So far we have dealt with $\bZ_2$ involutions that act by at most a
sign on the GLSM coordinates. This is enough for treating the examples
in the main text, but it is not the most general class of possible
$\bZ_2$ involutions one may consider. In this section we would like to
briefly discuss the extension of the ideas of the previous section to
involution exchanging GLSM coordinates. Namely, we allow actions of
the form:
\begin{align}
  \pi_2:(x_1,x_2,\ldots,x_n) \mapsto
  (x_{p_2(1)},x_{p_2(2)},\ldots,x_{p_2(n)})
\end{align}
where $p_2$ is an order 2 permutation of $\{1,2,\ldots,n\}$. Such
$\bZ_2$ involutions have been considered recently in the context of
F-theory model building \cite{Collinucci:2009uh,Blumenhagen:2009up}.

There are a few interesting subtleties that appear in this case.
Since now the permutation acts on the coordinates, it can act
non-trivially on the \v Cech complex we obtain out of each local
section. In principle, in order to compute equivariant line bundle
cohomology in this case we would need to define this action on the \v
Cech complex carefully, and obtain in this way the induced
representation on the \v Cech cohomology for each
chamber.\footnote{Here we are using ideas and terminology from the
  chamber algorithm for computing line bundle cohomology. We refer the
  reader to \cite{CLS:ToricVarieties,Blumenhagen:2010pv,Cvetic:2010rq}
  for reviews of the relevant concepts.} There is nevertheless a
simple shortcut we can use in order to avoid having to do this. Notice
that, since each local section in the same \v Cech chamber gives rise
to the same \v Cech complex, for each chamber we have a direct product
structure for the representation of the $\bZ_2$ action. Namely, when
computing the character of $\pi_2$ on a particular chamber $C$, we can
write:
\begin{align}
  \label{eq:split-representation}
  \Tr_C(\pi_2) = \Tr_{\check{c}}(\pi_2)\cdot \Tr_M(\pi_2)
\end{align}
where $\check{c}$ is the cohomology group coming from each section in
the chamber, and $M$ is the space of sections in the chamber. The
second term $\Tr_M(\pi_2)$ denotes the trace of the $\bZ_2$ action on
the space of local sections in the chamber under consideration. Notice
that we only need to consider chambers mapped to themselves under the
$\bZ_2$ action. If a chamber is not invariant the induced
representation on the local sections always acts as an exchange of
sections, i.e. a matrix of the form:
\begin{align}
  M = \begin{pmatrix}0&\pm 1\\\pm 1 & 0\end{pmatrix}
\end{align}
which has zero trace, and thus does not contribute to the equivariant
index.

Since we are taking $\Tr_{\check{c}}(\pi_2)$ to depend just on the
structure of the \v Cech complex in the chamber, we can use index
formulas to determine it in a few simple cases, and then use this for
obtaining the general result. Let us illustrate how this works in a
simple example. Consider the third del Pezzo surface $dP_3$, described
by the GLSM data in table~\ref{table:dP3} above, and take the
involution given by:
\begin{align}
  \label{eq:dP3-permutation}
  \pi_2: (x_1,x_2,x_3,x_4,x_5,x_6) \mapsto (x_2,x_1,x_3,x_4,x_6,x_5).
\end{align}

We are interested in computing line bundle cohomology for an
equivariant bundle of the form $\cO(m,n,p,p)$. We have chosen this
particular form in order for the line bundle to map to itself under
the involution $\pi_2$. It is a simple calculation to show that the
fixed point locus of~\eqref{eq:dP3-permutation} is given by the curve
$x_1x_5 - x_2x_6=0$ and the two points $\{x_1x_5+x_2x_6=0,x_4=0\}$,
$\{x_1x_5+x_2x_6=0,x_3=0\}$. The resulting Lefschetz equivariant index
is then given by:
\begin{align}
  \label{eq:permutation-index}
  \chi^{\pi_2}(dP_3,\cO(m,n,p,p)) = \frac{1}{2}(n+1) +
  \frac{(-1)^m}{4}(1+(-1)^n).
\end{align}
As an example, consider the line bundle $\cO(-2,0,0,0)$. The only
non-vanishing contribution from a symmetric chamber comes from an
element in the chamber $(x_1x_2)^{-1}$, understood as a term in the
power-set of the Stanley-Reisner ideal
\cite{Blumenhagen:2010pv,Jow:2010,Rahn:2010fm}, and in particular from
the single local section $x_5x_6/x_1x_2$. In the notation
of~\eqref{eq:split-representation} we thus have that $\Tr_M(\pi_2)=1$,
and since this is a contribution to $H^1(dP_3,\cO(-2,0,0,0))$, we have
that:
\begin{align}
  \chi^{\pi_2}(dP_3,\cO(-2,0,0,0)) = - \Tr_{\check{c}}(\pi_2) \cdot 1
  = - \Tr_{\check{c}}(\pi_2)
\end{align}
with $\check{c}$ being in this case the $(x_1x_2)^{-1}$ chamber. We
can now compute the equivariant index independently using
formula~\eqref{eq:permutation-index}, and from there we obtain:
\begin{align}
  \label{eq:dP3-Tr(12)}
  \Tr_{\check{c}}(\pi_2) = -1.
\end{align}
The trace over the other contributing chambers can then be computed
similarly. Knowledge of the result for all contributing chambers in
the problem then allows us to compute equivariant line bundle
cohomology for any line bundle
using~\eqref{eq:split-representation}. For instance, the character
$\chi_{\pi_2}(H^1(dP_3,\cO(-2,2,0,0)))$ receives contributions only
the trace over the $(x_1x_2)^{-1}$ chamber. The space of contributing
local sections in this chamber is generated by:
\begin{align}
  \begin{matrix}
    \frac{x_4^2x_5x_6}{x_1x_2}\quad&\quad\frac{x_3x_4x_5^2x_6}{x_1x_2^2}\quad&\quad
    \frac{x_3^2x_5^3x_6}{x_1x_2^3}\\
    \frac{x_3x_4x_5x_6^2}{x_1^2x_2}\quad&\quad\frac{x_3^2x_5^6x_6^2}{x_1^2x_2^2}\quad&\quad \frac{x_3^2x_5x_6^3}{x_1^3x_2}
  \end{matrix}
\end{align}
From here we find that $\Tr_M(\pi_2)=2$ (since there are 2 invariant
local sections), and using~\eqref{eq:dP3-Tr(12)} we then find that:
\begin{align}
  \chi_{\pi_2}(H^1(dP_3,\cO(-2,2,0,0))) = -2.
\end{align}
As a simple check, this result agrees with the one obtained from the
Lefschetz index~\eqref{eq:permutation-index}. Since
$\dim(H^1(dP_3,\cO(-2,2,0,0)))=8$, we have that
\begin{align}
  \begin{split}
    h^1_+(dP_3,\cO(-2,2,0,0))&=3\\
    h^1_-(dP_3,\cO(-2,2,0,0))&=5.
  \end{split}
\end{align}

\section{Factorization: a geometric viewpoint}
\label{sec:geometric-viewpoint}

In section \ref{sec:elliptic-fibrations}, arguments for the
factorization of intersection forms on certain manifolds were presented
from an algebraic viewpoint. Specifically, given the Stanley-Reisner
ideal of a $d$-dimensional ambient space toric variety $\cA$ along
with the data encoding linear equivalence, one can calculate the
intersection ring explicitly to determine whether or not it
factorizes. In the case where generators of the Stanley-Reisner ideal
of some $k$-dimensional toric subvariety $\cB\subset \cA$ are also
generators of the Stanley-Reisner ideal of $\cA$ and linear
equivalence of divisors in $\cB$ is preserved in $\cA$, then divisors
in $\cB$ pulled back to the ambient space can appear at most $k$ times
in non-vanishing monomials in the intersection $d$-form of $\cA$.

While the algebraic viewpoint makes everything explicit, one of the
well-known virtues of toric varieties is that they are amenable to
powerful combinatorial methods of analysis. For example, one set of
lattice data which can be used to define a toric variety is a
\emph{complete fan}, which is a set of \emph{strongly convex rational
  polyhedral cones}, each of which corresponds to an affine patch on
the toric variety, along with some consistency conditions which the
cones must satisfy. Each divisor corresponds to a ray in the lattice,
and a set of divisors $\{D_{i_1},\dots,D_{i_k}\}$ (up to linear
equivalence) can simultaneously vanish if and only if the
corresponding rays are in some cone in the fan. If there is no such
cone, this is the combinatorial description of the statement that
$x_{i_1}\dots x_{i_k}$ is in the Stanley-Reisner ideal.

Though we worked primarily with GLSM and algebraic data for
discussions in the main text, in this appendix we would like to
discuss briefly the combinatorial viewpoint on manifolds whose
intersection form factorizes. Those readers familiar with toric
geometry can probably already anticipate the answer: since the
elliptic Calabi-Yau manifolds in the main text were constructed in a
toric ambient space $\cA$ by essentially taking the GLSM data
corresponding to some $d$-dimensional toric variety $\cB$ and
augmenting it with a copy of $\bP_{231}$, it is possible to see
explicitly both the $d$-dimensional lattice data of $\cB$ and the
two-dimensional lattice data of $\bP_{231}$ reflected in the
$(d+2)$-dimensional lattice data of $\cA$.

Before going into the general discussion, let us present an example to
motivate it. We use as our example the GLSM data given in table
\ref{table:T2overFn} for a toric variety whose Calabi-Yau hypersurface
is an elliptic fibration over the Hirzebruch surface $\bF_n$. The
point matrix
\begin{align}
  v_{MN} = \left(\begin{array}{cccc|ccc}
    1 & -1 & 0 & 0 \, &\, 0 & 0 & 0\\
    0 & -n & 1 & -1\, &\, 0 & 0 & 0\\
    \hline
    -2 & -2 & -2 & -2\, &\, 1 & 0 & -2\\
    -3 & -3 & -3 & -3\, &\, 0 & 1 & -3
    \end{array}\right)
\end{align}
satisfies the relation $v_{MN}Q^a_N=0\,\,\,\,\forall a$, as required
by the theory of toric varieties, and the $Q^a$ vectors are just
$M,N,O,P$ in the GLSM data.  It is straightforward to check that this
is indeed a solution, where the columns give the points in the
four-dimensional $N$ lattice corresponding to $s$,$t$,$u$,$v$,$x$,$y$
and $z$, respectively.  Moreover, the lattice data corresponding to
both $\bF_n$ and $\bP_{231}$ are present in the upper left quadrant
and the lower right quadrant, respectively.  In fact, it can be seen
by working out the linear algebra that this solution does not depend
explicitly on the fan data corresponding to $\cB$, but instead just on
the fact that the point matrix of $\cB$ annihilates the subset of the
GLSM data associated with $\cB$.

This fact suggests a generalization of the solution for
$\cB=\bF_n$. Let $\cB$ be a $d$-dimensional toric variety with $k$
homogeneous coordinates, which therefore has $(k-d)$ GLSM relations.
Taking $A$ to be the $(d\times k)$ point matrix of $\cB$, $B$ and $C$
to be precisely the lower left and lower right quadrants in the
$\bF_n$ solution, and $\textbf{0}$ to be a $(d\times 3)$ matrix of
zeroes, then
\begin{align}
v_{M N} = \left( \begin{array}{c|c}
	A & \textbf{0} \\ \hline
	B & C
      \end{array} \right)
\end{align}
is the point matrix of a $(d+2)$-dimensional toric variety with the
point matrices of $\cB$ and $\bP_{231}$ explicit in the upper left and
lower right quadrants. It has $(k+3)$ homogeneous coordinates and
$(k-d+1)$ relations given by
\begin{equation}
  \tilde Q^a = (Q^a_i, 2\sum_i Q^a_i, 3\sum_i Q^a_i,0)
  \qquad \text{and} \qquad
  \tilde Q^{a+1} = (\overbrace{0,\dots,0}^{{}\text{k times}},2,3,1),
\end{equation}
where $Q^a_i$ are the $k$-dimensional charge vectors of $\cB$. The
fact that the point matrix annihilates these charge vectors relies
heavily on the fact that the Calabi-Yau condition determined the last
three components of $\tilde Q^a$.  All $\bP_{231}$ fibrations in the
main text have GLSM data of this form.

From this example and our general solution, we see that the
$k$-dimensional subspace $N_\cB$ where the lattice data of $\cB$ lives
is $(x_1, \dots, x_k, -2,-3)\in \bR^{d+2}$ and the $2$-dimensional
subspace $N_{231}$ where the lattice data of $\bP_{231}$ lives is $(0,
\dots, 0, x_{d+1},x_{d+2})\in \bR^{d+2}$.  An interesting observation
from the Hirzebruch example and also from the general solution is that
the point corresponding to the $z$ coordinate is precisely at the
origin of $N_\cB$, i.e. at $(0, \dots, 0, -2, -3)$. If $\cB$ is a
toric variety defined by a $*$-triangulation of a polytope $B \subset
N_\cB$ (whose origin is the $z$ point in the $(k+2)$-dimensional
lattice $N$), then every simplex in the triangulation necessarily
contains the $z$ point. Since linear equivalence of divisors in $\cB$
is preserved amongst their pullbacks in $\cA$, any non-zero
intersection of $k+1$ divisors in $\cA$ whose points are in $N_\cB$
must necessarily involve $D_z$. This is equivalent to factorization of
the intersection $(k+2)$-form on $\cA$, given our assumption above
that the $Div(\cA)$ generator corresponding to $\tilde Q^{a+1}$ is the
only one which is not also a generator of $Div(\cB)$.

\bibliographystyle{JHEP}
\bibliography{refs}

\end{document}